\documentstyle[amsfonts,aps,epsf,amsbsy,twocolumn]{revtex}

\newcommand{\BiInfinity}			{ \stackrel{\leftrightarrow} {S} }
\newcommand{\biinfinity}			{ \stackrel{\leftrightarrow} {s} }
\newcommand{\Past}        		{ \stackrel{\leftarrow} {S} }
\newcommand{\past}        		{ {\stackrel{\leftarrow} {s}} }
\newcommand{\Future}      		{ \stackrel{\rightarrow}{S} }
\newcommand{\future}      		{ \stackrel{\rightarrow}{s} }
\newcommand{\PastPrime}			{ {\stackrel{\leftarrow}{s}}^\prime }

\newcommand{\PastDblPrime}		{ {\stackrel{\leftarrow}{s}}^{\prime\prime} }

\newcommand{\pastL}       		{ {\stackrel{\leftarrow} {s}}^L }
\newcommand{\PastL}				{ {\stackrel{\leftarrow} {S}}^L }
\newcommand{\PastLt}			{ {\stackrel{\leftarrow} {S}}_t^L }
\newcommand{\PastLLessOne}		{ {\stackrel{\leftarrow} {S}}^{L-1} }
\newcommand{\futureL}     		{ {\stackrel{\rightarrow}{s}}^L }
\newcommand{\FutureL}			{ {\stackrel{\rightarrow}{S}}^L }
\newcommand{\FutureLt}			{ {\stackrel{\rightarrow}{S}}_t^L }
\newcommand{\FutureLLessOne}		{ {\stackrel{\rightarrow}{S}}^{L-1} }

\newcommand{\AllPasts}			{ { \stackrel{\leftarrow} {\rm {\bf S}} } }
\newcommand{\AllFutures}			{ \stackrel{\rightarrow} {\rm {\bf S}} }
\newcommand{\FutureSet}			{ \stackrel{\rightarrow}{\bf S}}
\newcommand{\CausalState}		{ {\cal S} }
\newcommand{\causalstate}		{ \sigma }
\newcommand{\CausalStateSet}		{ \boldsymbol{\CausalState} }
\newcommand{\AlternateState}		{ {\cal R} }
\newcommand{\alternatestate}		{ \rho }
\newcommand{\AlternateStateSet}	{ \boldsymbol{\AlternateState} }
\newcommand{\PrescientState}		{ \hat{\AlternateState} }
\newcommand{\prescientstate}		{ \hat{\alternatestate} }
\newcommand{\PrescientStateSet}	{ \boldsymbol{\PrescientState}}
\newcommand{\CausalEquivalence}	{ {\sim}_{\epsilon} }

\newcommand{\NextObservable}		{ {\stackrel{\rightarrow} {S}}^1 }
\newcommand{\LastObservable}		{ {\stackrel{\leftarrow} {S}}^1 }
\newcommand{\Prob}				{ {\rm P} }
\newcommand{\ProbAnd}			{ {,\;} }
\newcommand{\LLimit}			{ L \rightarrow \infty }
\newcommand{\Cmu}				{ C_\mu }
\newcommand{\hmu}				{ h_\mu }
\newcommand{\EE}				{ {\bf E}}
\newtheorem{theorem}{Theorem}
\newtheorem{lemma}{Lemma}
\newtheorem{corollary}{Corollary}
\newtheorem{definition}{Definition}
\newtheorem{proposition}{Proposition}

\begin{document}
\draft
% *******************Title Page************************
\title{Computational Mechanics:\\
Pattern and Prediction, Structure and Simplicity}

\author{Cosma Rohilla Shalizi\thanks{Permanent address: Physics
Department, University of Wisconsin, Madison, WI 53706}
and James P. Crutchfield}
\address{Santa Fe Institute, 1399 Hyde Park Road, Santa Fe, NM 87501\\
Electronic addresses: \{shalizi,chaos\}@santafe.edu}

\date{\today}
\maketitle

\bibliographystyle{unsrt}

%*************************Abstract*********************
\begin{abstract}
Computational mechanics, an approach to structural complexity, defines a
process's causal states and gives a procedure for finding them.  We show that
the causal-state representation---an $\epsilon$-machine---is the minimal one
consistent with accurate prediction.  We establish several results on
$\epsilon$-machine optimality and uniqueness and on how $\epsilon$-machines
compare to alternative representations.  Further results relate measures of
randomness and structural complexity obtained from $\epsilon$-machines to those
from ergodic and information theories.

\begin{center}
Santa Fe Institute Working Paper 99-07-044\\
{\bf Keywords}: complexity, computation, entropy,\\
information, pattern, statistical mechanics.\\
{\em Running Head}: Computational Mechanics
\end{center}
\end{abstract}
%******************************************************

\pacs{02.50.Wp, 05.45, 05.65+b, 89.70.+c}

%\begin{quotation}
%\hspace{-0.5in}One man's rigor is another man's mortis.\\
%\\
%---C. Bohren and B. Albrecht in
%{\em Atmospheric Thermodynamics} \cite{Bohren-Albrecht}.
%\end{quotation}

\vspace{-0.25in}
\tableofcontents

%************************* INTRODUCTION *************************

\section{Introduction}

Organized matter is ubiquitous in the natural world, but the branch of physics
which ought to handle it---statistical mechanics---lacks a coherent, principled
way of describing, quantifying, and detecting the many different kinds of
structure nature exhibits.  Statistical mechanics has good measures of disorder
in thermodynamic entropy and in related quantities, such as the free energies.
When augmented with theories of critical phenomena \cite{Yeomans} and pattern
formation \cite{Manneville-dissipative-structures}, it also has an extremely successful approach to
analyzing patterns formed through symmetry breaking, both in equilibrium
\cite{Chaikin-Lubensky} and, more recently, outside it \cite{Cross-Hohenberg}.  Unfortunately,
these successes involve many {\it ad hoc} procedures---such as guessing
relevant order parameters, identifying small parameters for perturbation
expansion, and choosing appropriate function bases for spatial decomposition.
It is far from clear that the present methods can be extended to handle all the
many kinds of organization encountered in nature, especially those produced by
biological processes.

{\it Computational mechanics} \cite{Calculi-of-emergence} is an approach that lets us
directly address the issues of pattern, structure, and organization. While
keeping concepts and mathematical tools already familiar from statistical
mechanics, it is distinct from the latter and complementary to it. In essence,
from either empirical data or from a probabilistic description of behavior, it
shows how to infer a model of the hidden process that generated the observed
behavior. This representation---the $\epsilon$-machine---captures the patterns
and regularities in the observations in a way that reflects the causal
structure of the process. Usefully, with this model in hand, one can extrapolate
beyond the original observational data to make predictions of future behavior.
Moreover, in a well defined sense that is the subject of the following, the
$\epsilon$-machine is the unique maximally efficient model of the observed
data-generating process.

$\epsilon$-Machines themselves reveal, in a very direct way, how information is
stored in the process, and how that stored information is transformed by new
inputs and by the passage of time.  This, and not using computers for
simulations and numerical calculations, is what makes computational mechanics
``computational'', in the sense of ``computation theoretic''.

The basic ideas of computational mechanics were introduced a decade ago
\cite{Inferring-stat-compl}.  Since then they have been used to analyze dynamical systems
\cite{Computation-at-the-onset,Perry-Binder-finite-stat-compl}, cellular automata \cite{Comp-mech-of-CA-example}, hidden Markov models
\cite{Upper-thesis}, evolved spatial computation \cite{JPC-MM-PNAS},
stochastic resonance \cite{Witt-et-al-1997}, globally coupled maps \cite{Delgado-collective-induced}, and
the dripping faucet experiment \cite{Goncalves-dripping-faucet}.  Despite this record of
successful application, there has been some uncertainty about the mathematical
foundations of the subject.  In particular, while it seemed evident from
construction that an $\epsilon$-machine captured the patterns inherent in a
process and did so in a minimal way, no explicit proof of this was published.
Moreover, there was no proof that, if the $\epsilon$-machine was optimal in
this way, it was the {\it unique} optimal representation of a process.  These
little-needed gaps have now been filled.  Subject to some (reasonable)
restrictions on the statistical character of a process, we prove that the
$\epsilon$-machine is indeed the unique optimal causal model.  The rigorous
proof of these results is the main burden of this paper.  We gave preliminary
versions of the optimality results---but not the uniqueness theorem, which is
new here---in Ref.~\cite{TDCS}.

The outline of the exposition is as follows.  We begin by showing how
computational mechanics relates to other approaches to pattern, randomness, and
causality.  The upshot of this is to focus our attention on {\it patterns
within a statistical ensemble} and their possible representations.  Using ideas
from information theory, we state a quantitative version of Occam's Razor for
such representations.  At that point we define {\it causal states}
\cite{Inferring-stat-compl}, equivalence classes of behaviors, and the structure of
transitions between causal states---the $\epsilon$-machine.  We then show that
the causal states are ideal from the point of view of Occam's Razor, being the
simplest way of attaining the maximum possible predictive power.  Moreover, we
show that the causal states are uniquely optimal. This combination allows us
to prove a number of other, related optimality results about
$\epsilon$-machines. We examine the assumptions made in deriving these
optimality results, and we note that several of them can be lifted without
unduly upsetting the theorems. We also establish bounds on a process's
{\em intrinsic computation} as revealed by $\epsilon$-machines and by
quantities in information and ergodic theories. Finally, we close by reviewing
what has been shown and what seem like promising directions for further work
on the mathematical foundations of computational mechanics.

A series of appendices provide supplemental material on information theory,
equivalence relations and classes, $\epsilon$-machines for time-reversed
processes, semi-group theory, and connections and distinctions between
computational mechanics and other fields.

To set the stage for the mathematics to follow and to motivate the assumptions
used there, we begin now by reviewing prior work on pattern, randomness, and
causality. We urge the reader interested only in the mathematical development
to skip directly to Sec. \ref{Desiderata}---a synopsis of the central
assumptions of computational mechanics---and continue from there.

\section{Patterns}
\label{Patterns}

To introduce our approach to---and even to argue that {\em some} approach is
necessary for---discovering and describing patterns in nature we begin by
quoting Jorge Luis Borges:
\begin{quotation}
These ambiguities, redundancies, and deficiencies recall those attributed by
Dr.~Franz Kuhn to a certain Chinese encyclopedia entitled {\it Celestial
Emporium of Benevolent Knowledge}.  On those remote pages it is written that
animals are divided into (a) those that belong to the Emperor, (b) embalmed
ones, (c) those that are trained, (d) suckling pigs, (e) mermaids, (f) fabulous
ones, (g) stray dogs, (h) those that are included in this classification, (i)
those that tremble as if they were mad, (j) innumerable ones, (k) those drawn
with a very fine camel's hair brush, (l) others, (m) those that have just
broken a flower vase, (n) those that resemble flies from a distance.\\
\\
---J. L. Borges, ``The Analytical Language of John Wilkins'', in
   Ref.~\cite[p.~103]{Borges-inquisitions}; see also discussion in Ref.~\cite{JPC-semantics}.
\end{quotation}

The passage illustrates the profound gulf between patterns, and classifications
derived from patterns, that are appropriate to the world and help us to
understand it and those patterns which, while perhaps just as legitimate as
prosaic regularities, are not at all informative.  What makes the
{\it Celestial Emporium's} scheme
inherently unsatisfactory, and not just strange, is that it tells us nothing
about animals.  We want to find patterns in a process that ``divide it at the
joints, as nature directs, not breaking any limbs in half as a bad carver
might'' \cite[Sec.~265D]{Phaedrus}.

Computational mechanics is not directly concerned with pattern formation {\it
per se} \cite{Cross-Hohenberg}; though we suspect it will ultimately be useful in that
domain.  Nor is it concerned with pattern recognition as a practical matter as
found in, say, neuropsychology \cite{Luria-working-brain}, psychophysics \cite{Graham-pattern-analyzers},
cognitive ethology \cite{Shettleworth}, computer engineering \cite{Tou-and-Gonzalez}, and signal
and image processing \cite{Banks-processing,Lim-two-d-processing}.  Instead, it is concerned with the
questions of {\it what patterns are} and {\it how patterns should be
represented}.  One way to highlight the difference is to call this pattern {\em
discovery}, rather than pattern {\it recognition}.

The bulk of the intellectual discourse on what patterns are has been
philosophical.  One distinct subset has been conducted under the broad rubric
of mathematical logic.  Within this there are approaches, on the one hand, that
draw on (highly) abstract algebra and the theory of relations; on the other,
that approach patterns via 
the theory of algorithms and effective procedures.

The general idea, in both approaches, is that some object ${\cal O}$ has a
pattern ${\cal P}$---${\cal O}$ has a pattern ``represented'', ``described'',
``captured'', and so on by ${\cal P}$---if and only if we can use ${\cal P}$ to
predict or compress ${\cal O}$.  Note that the ability to predict implies the
ability to compress, but not vice versa; here we stick to prediction.  The
algebraic and algorithmic strands differ mainly on how ${\cal P}$ itself should
be represented; that is, they differ in how it is expressed in the vocabulary
of some formal scheme.

We should emphasize here that ``pattern'' in this sense implies a kind of
regularity, structure, symmetry, organization, and so on.  In contrast,
ordinary usage sometimes accepts, for example, speaking about the ``pattern''
of pixels in a particular slice of between-channels video ``snow''; but we
prefer to speak of that as the {\it configuration} of pixels.

\subsection{Algebraic Patterns}

Although the problem of pattern discovery appears early, in Plato's {\it Meno}
\cite{Meno} for example, perhaps the first attempt to make the notion of
``pattern'' mathematically rigorous was that of Whitehead and Russell in
{\it Principia Mathematica}. They viewed pattern as a property, not of sets,
but of relations within or between sets, and accordingly they work out an
elaborate {\em relation-arithmetic} \cite[vol.~II, part IV]{Principia};
cf.~\cite[ch.~5--6]{Russell-intro-to-math-phil}. This starts by defining the {\em relation-number}
of a relation between two sets as the class of all the relations that are
equivalent to it under one-to-one, onto mappings of the two sets.
In this framework relations share a common pattern or structure if they
have the same relation-number. For instance, all square lattices have
similar structure since their elements share the same neighborhood relation;
as do all hexagonal lattices. Hexagonal and square lattices, however, exhibit
different patterns since they have non-isomorphic neighborhood relations---i.e.,
since they have different relation-numbers. (See also {\em recoding equivalence}
defined in Ref.~\cite{JPC-information-and-its-metric}.) Less work has been done on this than
they---especially Russell \cite{Russell-human-knowledge}---had hoped. This may be due in part
to a general lack of familiarity with Volume II of Ref.~\cite{Principia}.

A more recent attempt at developing an algebraic approach to patterns builds on
semi-group theory and its Krohn-Rhodes decomposition theorem.
Ref.~\cite{Rhodes-book} discusses a range of applications of this approach to
patterns. Along these lines, Rhodes and Nehaniv have tried to apply semi-group
complexity theory to biological evolution \cite{Nehaniv-Rhodes-evolution}. They suggest that
the complexity of a biological structure can be measured by the number of
subgroups in the decomposition of an automaton that describes the structure.

Yet another algebraic approach has been developed by Grenander and co-workers,
primarily for pattern recognition \cite{Grenander-elements}.  Essentially, this is a matter
of trying to invent a minimal set of {\em generators} and {\em bonds} for the
pattern in question.  Generators can adjoin each other, in a suitable
$n$-dimensional space, only if their bonds are compatible.  Each pair of
compatible bonds at once specifies a binary algebraic operation and an
observable element of the configuration built out of the generators.  (Our
construction in App.~\ref{SemiGroups}, linking an algebraic operation with
concatenations of strings, is analogous in a rough way.) Probabilities can be
attached to these bonds, leading in a natural way to a (Gibbsian) probability
distribution over entire configurations.  Grenander and his colleagues have
used these methods to characterize, {\it inter alia}, several
biological phenomena \cite{Grenander-hands,Grenander-potatoes}.

\subsection{Turing Mechanics: Patterns and Effective Procedures}

The other path to patterns follows the traditional exploration of the logical
foundations of mathematics, as articulated by Frege and Hilbert and pioneered
by Church, G\"odel, Post, Russell, Turing, and Whitehead.  A more recent and
relatively more popular approach goes back to Kolmogorov and Chaitin, who were
interested in the {\it exact} reproduction of an individual object
\cite{Kolmogorov-three-approaches,Chaitin-KC-complexity-cite,Kolmogorov-1983,Li-and-Vitanyi-1993}; in particular, their focus was discrete
symbol systems, rather than (say) real numbers or other mathematical objects.
The candidates for expressing the pattern ${\cal P}$ were universal Turing
machine (UTM) programs---specifically, the shortest UTM program that can
exactly produce the object ${\cal O}$.  This program's length is called ${\cal
O}$'s {\em Kolmogorov-Chaitin complexity}.  Note that any scheme---automaton,
grammar, or what-not---that is Turing equivalent and for which a notion of
``length'' is well defined will do as a representational scheme.  Since we can
convert from one such device to another---say, from a Post tag system
\cite{Minsky-computation} to a Turing machine---with only a finite description of the first
system, such constants are easily assimilated when measuring complexity in this
approach.

In particular, consider the first $n$ symbols ${\cal O}_{n}$ of ${\cal O}$ and
the shortest program ${\cal P}_{n}$ that produces them.  We ask, What happens
to the limit
\begin{equation}
\lim_{n \rightarrow \infty}{{|{\cal P}_{n}| \over n}} ~,
\end{equation}
where $|{\cal P}|$ is the length in bits of program ${\cal P}$? On the one
hand, if there is a fixed-length program ${\cal P}$ that generates arbitrarily
many digits of ${\cal O}$, then this limit vanishes.  Most of our interesting
numbers, rational or irrational---such as $\pi$, $e$, $\sqrt{2}$---are of this
sort.  These numbers are eminently compressible: the program ${\cal P}$ is the
compressed description, and so it captures the pattern obeyed by the sequence
describing ${\cal O}$.  If the limit goes to $1$, on the other hand, we have a
completely incompressible description and conclude, following Kolmogorov,
Chaitin, and others, that $\cal O$ is random
\cite{Kolmogorov-three-approaches,Chaitin-KC-complexity-cite,Kolmogorov-1983,Li-and-Vitanyi-1993,Martin-Lof,Levin-information-conservation}.  This conclusion is the
desired one: the Kolmogorov-Chaitin framework establishes, formally at least,
the randomness of an individual object without appeals to probabilistic
descriptions or to ensembles of reproducible events.  And it does so by
referring to a deterministic, algorithmic representation---the UTM.

There are many well-known difficulties with applying Kolmogorov complexity to
natural processes.  First, as a quantity, it is uncomputable in general, owing
to the halting problem \cite{Li-and-Vitanyi-1993}.  Second, it is maximal for random
sequences; this can be construed either as desirable, as just noted, or as a
failure to capture structure, depending on one's aims.  Third, it only applies
to a single sequence; again this is either good or bad.  Fourth, it makes no
allowance for noise or error, demanding exact reproduction.  Finally, $\lim_{n
\rightarrow \infty} |{\cal P}_{n}| / n$ can vanish, although the computational
resources needed to run the program, such as time and storage, grow without
bound.

None of these impediments have kept researchers from attempting to use
Kolmogorov-Chaitin complexity for practical tasks---such as measuring the
complexity of natural objects (e.g.~Ref.~\cite{Gurzadyan}), as a basis for
theories of inductive inference \cite{Solomonoff,Vitanyi-and-Li-1999}, and generally as a
means of capturing patterns \cite{Flake}.  As Rissanen \cite[p.~49]{Rissanen-SCiSI}
says, this is akin to ``learn[ing] the properties [of a data set] by writing
programs in the hope of finding short ones!''

Various of the difficulties just listed have been addressed by subsequent work.
Bennett's {\it logical depth} accounts for time resources \cite{Bennett-how-and-why}.  (In
fact, it is the time for the minimal-length program ${\cal P}$ to produce
$\cal O$.) Koppel's {\it sophistication} attempts to separate out the
``regularity'' portion of the program from the random or instance-specific
input data \cite{Koppel-1987,Koppel-Atlan}. Ultimately, these extensions and
generalizations remain in the UTM, exact-reproduction setting and so inherit
inherent uncomputability.

\subsection{Patterns with Error}

Motivated by these theoretical difficulties and practical concerns, an obvious
next step is to allow our pattern ${\cal P}$ some degree of approximation or
error, in exchange for shorter descriptions.  As a result, we lose perfect
reproduction of the original configuration from the pattern.  Given the
ubiquity of noise in nature, this is a small price to pay.  We might also say
that sometimes we are willing to accept small deviations from a regularity,
without really caring what the precise deviation is.  As pointed out in
Ref.~\cite{JPC-semantics}'s conclusion, this is certainly a prime motivation in
thermodynamic descriptions, in which we explicitly throw away, and have no
interest in, vast amounts of microscopic detail in order to find a workable
description of macroscopic observations.

Some interesting philosophical work on patterns-with-error has been done by
Dennett, with reference not just to questions about the nature of
patterns and their emergence but
also to psychology \cite{Dennett-real-patterns}.  The intuition is that truly random
processes can be modeled very simply---``to model coin-tossing, toss a coin.''
Any prediction scheme that is more accurate than assuming complete independence
{\it ipso facto} captures a pattern in the data.  There is thus a spectrum of
potential pattern-capturers ranging from the assumption of pure noise to
the exact reproduction of the data, if that is possible.  Dennett notes that
there is generally a trade-off between the simplicity of a predictor and its
accuracy, and he plausibly describes emergent phenomena \cite{Anything-ever-new,Holland-emergence}
as patterns that allow for a large reduction in complexity for only a small
reduction in accuracy.  Of course, Dennett was by no means the first to
consider predictive schemes that tolerate error and noise; we discuss some of
the earlier work in App.~\ref{compare-and-contrast}. However, to our knowledge,
he was the first to have made such predictors a central part of an explicit
account of {\it what patterns are}.  It must be noted that this account lacks
the mathematical detail of the other approaches we have considered so far, and
that it relies on the inexact prediction of a single configuration. In fact,
it relies on exact predictors that are ``fuzzed up'' by noise. The introduction
of noise, however, brings in probabilities, and their natural setting is in
ensembles. It is in that setting that the ideas we share with Dennett can
receive a proper quantitative treatment.

\subsection{Randomness: The Anti-Pattern?}

We should at this point say a bit about the relations between {\it randomness},
{\it complexity}, and {\it structure}, at least as we use those words. Ignoring
some foundational issues, randomness is actually rather well understood and
well handled by classical tools introduced by Boltzmann \cite{Boltzmann-gas-theory}; Fisher,
Neyman, and Pearson \cite{Cramer}; Kolmogorov \cite{Kolmogorov-three-approaches}; and Shannon
\cite{Shannon-1948}, among others.  One tradition in the study of complexity in fact
identifies complexity with randomness and, as we have just seen, this is
useful for some purposes.  As these purposes are {\it not} those
of analyzing patterns in processes and in real-world data, however, they are
not ours.  Randomness simply does not correspond to a notion of pattern or
structure at all and, by implication, neither Kolmogorov-Chaitin complexity nor
any of its spawn measure pattern.

Nonetheless, some approaches to complexity conflate ``structure'' with the
opposite of randomness, as conventionally understood and measured in physics by
thermodynamic entropy or a related quantity, such as Shannon entropy.  In
effect, structure is defined as ``one minus disorder''.  In contrast, we see
pattern---structure, organization, regularity, and so on---as describing a
coordinate ``orthogonal'' to a process's degree of randomness. That is,
complexity (in our sense) and randomness each capture a useful property
necessary to describe how a process manipulates information. This
complementarity is even codified by the complexity-entropy diagrams
introduced in Ref.~\cite{Inferring-stat-compl}.  It should be clear now that when we use the
word ``complexity'' we mean ``degrees'' of pattern, not degrees of randomness.

\subsection{Causation}
\label{Causation}

We want our representations of patterns in dynamical processes to be
causal---to say how one state of affairs leads to or produces another.
Although a key property, causality enters our development only in an
extremely weak sense,
the weakest one can use mathematically, which is Hume's \cite{Hume-treatise}: one
class of event causes another if the latter always follows the former; the
effect invariably succeeds the cause.  As good indeterminists, in the following
we replace this invariant-succession notion of causality with a more
probabilistic one, substituting a homogeneous distribution of successors for
the solitary invariable successor.  (A precise statement appears in
Sec.~\ref{DefnCausalStatesEMs}'s definition of {\it causal states}.) This
approach results in a purely phenomenological statement of causality, and so it
is amenable to experimentation in ways that stronger notions of
causality---e.g., ~that of Ref.~\cite{Bunge}---are not.
Ref.~\cite{Salmon-1984} independently reaches a concept of causality
essentially the same ours via philosophical arguments.

\subsection{Synopsis of Pattern}
\label{Desiderata}

In line with these observations, the ideal, synthesizing approach to patterns
would be at once:

\begin{enumerate}
\item {\em Algebraic}, giving us an explicit breakdown or decomposition of the
pattern into its parts;
\item {\em Computational}, showing how the process stores and uses information;
\item {\em Calculable}, analytically or by systematic approximation;
\item {\em Causal}, telling us how instances of the pattern are actually
produced; and
\item {\em Naturally stochastic}, not merely tolerant of noise but explicitly
formulated in terms of ensembles.
\end{enumerate}
This mix is precisely the brew we claim, in all modesty, to have on tap.

\section[Paddling around Occam's Pool]{Patterns in Ensembles:\\ Paddling
around Occam's Pool}

Here a pattern ${\cal P}$ is something knowledge of which lets us predict, at
better than chance rates, if possible, the future of sequences drawn from an
ensemble ${\cal O}$: ${\cal P}$ has to be statistically accurate and confer
some leverage or advantage as well.  Let's fix some notation and state the
assumptions that will later let us prove the basic results.

\subsection{Hidden Processes}
\label{HiddenProcesses}

We restrict ourselves to discrete-valued, discrete-time stationary stochastic
processes.  (See Sec.~\ref{ThingsThatAreNotYetTheorems} for discussion of these
assumptions.)  Intuitively, such processes are sequences of random variables
$S_i$, the values of which are drawn from a countable set ${\cal A}$.  We let
$i$ range over all the integers, and so get a bi-infinite sequence
\begin{equation}
\BiInfinity = \ldots S_{-1} S_0 S_1 \ldots ~.
\end{equation}
In fact, we define a process in terms of the distribution of such sequences;
cf.~Ref.~\cite{Billingsley-ergodic-theory-and-info}.

\begin{definition}[A Process]
Let ${\cal A}$ be a countable set.  Let $\Omega = {\cal A}^{\bf {\rm Z}}$ be
the set of bi-infinite sequences composed from ${\cal A}$, $T_i : \Omega
\mapsto {\cal A}$ be the function that returns the ${i}^{th}$ element ${s}_{i}$
of a bi-infinite sequence $\omega \in \Omega$, and ${\cal F}$ the field of
cylinder sets of $\Omega$.  Adding a probability measure $\Prob$ gives us a
probability space $(\Omega, {\cal F}, \Prob)$, with an associated random
variable $\BiInfinity$.  A {\rm process} is a sequence of random variables
${S}_{i} = {T}_{i}(\BiInfinity), i \in {\Bbb Z}$.
\label{AProcess}
\end{definition}
\addcontentsline{toc}{subsection}{\numberline{}Processes Defined}
Here, and throughout, we follow the convention of using capital letters to
denote random variables and lower-case letters their particular values.

It follows from Def.~\ref{AProcess} that there are well defined probability
distributions for sequences of every finite length.  Let $\FutureLt$ be the
sequence of $S_t, S_{t+1}, \ldots, S_{t+L-1}$ of $L$ random variables beginning
at $S_t$.  $\Future_t^0 \equiv \lambda$, the null sequence.  Likewise,
$\PastLt$ denotes the sequence of $L$ random variables going up to $S_t$, but
not including it; $\PastLt = \Future_{t-L}^L$.  Both $\FutureLt$ and $\PastLt$
take values from $s^L \in {\cal A}^L$.  Similarly, $\Future_t$ and $\Past_t$
are the semi-infinite sequences starting from and stopping at $t$ and taking
values $\future$ and $\past$, respectively.

Intuitively, we can imagine starting with distributions for finite-length
sequences and extending them gradually in both directions, until the infinite
sequence is reached as a limit.  While this can be a useful picture to have in
mind, defining a process in this way raises some subtle measure-theoretic
issues, such as how finite-dimensional distributions limit on an
infinite-dimensional one \cite[ch.~7]{Billingsley-probability-and-measure}. To
avoid these we start with the infinite-dimensional distribution.

\begin{definition}[Stationarity]
A process $S_i$ is stationary if and only if
\begin{equation}
\Prob(\FutureLt = s^L) = \Prob(\Future_0^L = s^L) ~,
\end{equation}
for all $t \in {\Bbb Z}$, $L \in {\Bbb Z}^{+}$, and all $s^L \in {\cal A}^L$.
\end{definition}
\addcontentsline{toc}{subsection}{\numberline{}Stationarity}
In other words, a stationary process is one that is time-translation invariant.
Consequently, $\Prob(\Future_t = \future) = \Prob(\Future_0 = \future)$ and
$\Prob(\Past_t = \past) = \Prob(\Past_0 = \past)$, and so we drop the
subscripts from now on.

\subsection{The Pool}
\label{ThePool}

\begin{figure}
\epsfxsize=2.7in
\begin{center}
\leavevmode
\epsffile{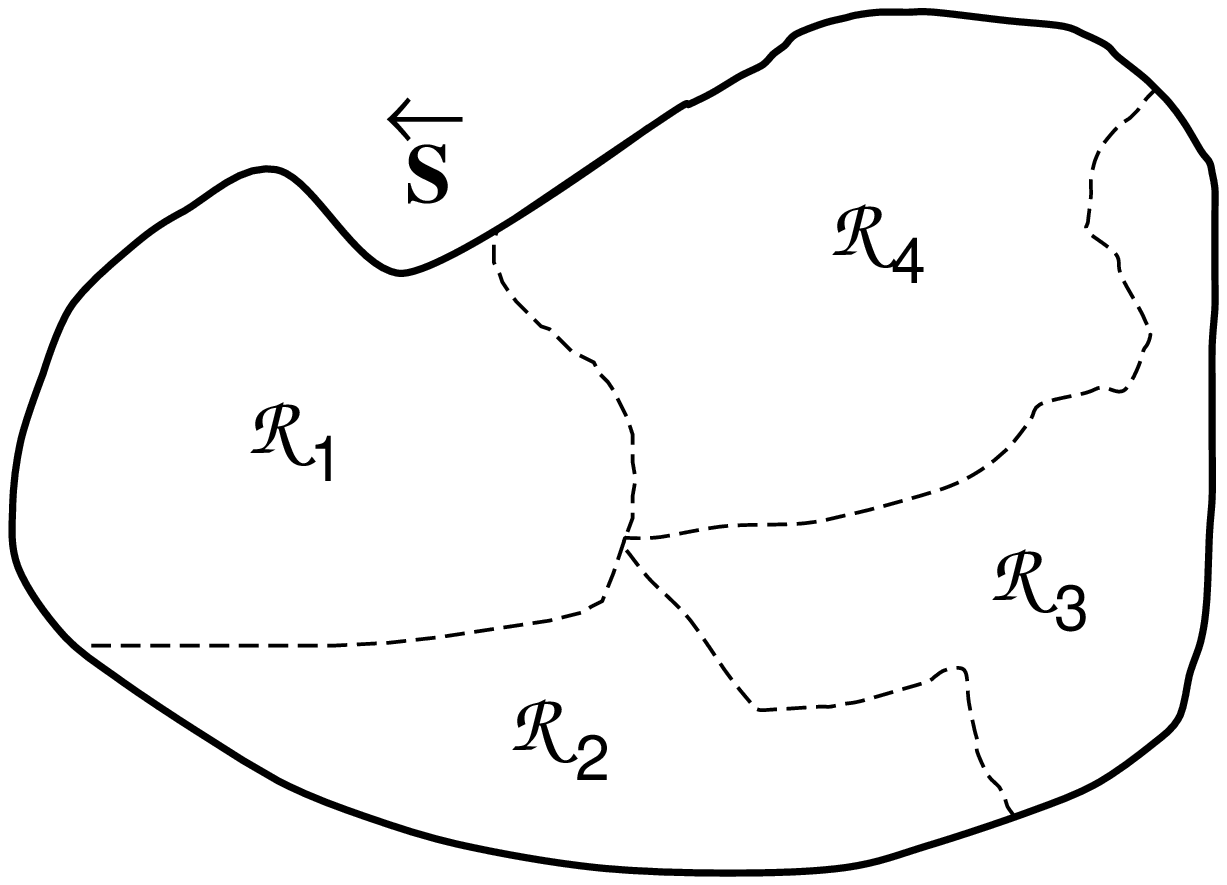}
\end{center}
\caption{A schematic picture of a partition of the set $\AllPasts$
  of all histories into some class of effective states:
  $\AlternateStateSet = \{ {\cal R}_i : i = 1, 2, 3, 4 \}$.  Note that the
  ${\cal R}_i$ need not form compact sets; we simply draw them
  that way for clarity. One should have in mind Cantor sets or other more
  pathological structures.
  }
\label{OccamsPool}
\end{figure}

Our goal is to predict all or part of $\Future$ using some function of some
part of $\Past$. We begin by taking the set $\AllPasts$ of all pasts and
partitioning it into mutually exclusive and jointly comprehensive subsets.
That is, we make a class $\AlternateStateSet$ of subsets of
pasts.\footnote{At several points our constructions require referring
to sets of sets. To help mark the distinction, we call the set
of sets of histories a {\em class}.}
(See Fig.~\ref{OccamsPool} for a schematic example.)  Each
$\alternatestate \in \AlternateStateSet$ will be called a {\it state} or an
{\it effective state}.  When the current history $\past$ is included in the
set $\alternatestate$, we will speak of the process being in state
$\alternatestate$.  Thus, we define a function from histories to effective
states:
\begin{equation}
\eta: \AllPasts \mapsto \AlternateStateSet ~.
\label{EtaDefn}
\end{equation}
A specific individual history $\past \in \AllPasts$ maps to a specific state
$\alternatestate \in \AlternateStateSet$; the random variable $\Past$ for the
past maps to the random variable $\AlternateState$ for the effective states.
It makes little difference whether we think of $\eta$ as being a function from
a history to a subset of histories or a function from a history to the
{\it label} of that subset. Each interpretation is convenient at different
times, and we will use both.

Note that we could use {\it any} function defined on $\AllPasts$ to partition
that set, by assigning to the same $\alternatestate$ all the histories $\past$
on which the function takes the same value.  Similarly, any equivalence
relation on $\AllPasts$ partitions it.  (See
App.~\ref{ReviewEquivalenceRelation} for more on equivalence relations.)  Due
to the way we defined a process's distribution, each effective state has a well
defined distribution of futures, though not necessarily a unique
one.\footnote{This is not necessarily true if $\eta$ is sufficiently
pathological.  To paraphrase Ref.~\cite{Schutz-geometrical}, readers should
assume that all our functions are sufficiently tame, measure-theoretically,
that whatever induced distributions we invoke will exist.}  Specifying the
effective state thus amounts to making a prediction about the process's
future. All the histories belonging to a given effective state are treated as
{\it equivalent for purposes of predicting the future}.  (In this way, the
framework formally incorporates traditional methods of time-series analysis;
see App.~\ref{TimeSeriesModeling}.)

We call the collection of all partitions $\AlternateStateSet$ of the set of
histories $\AllPasts$ {\it Occam's pool}.

\subsection{A Little Information Theory}

Since the bulk of the following development will be consumed with notions and
results from information theory \cite{Shannon-1948}, we now review several highlights
briefly, for the benefit of readers unfamiliar with the theory and to fix
notation.  Appendix \ref{InfoTheoryFormulae} lists a number of useful
information-theoretic formul{\ae}, which get called upon in our proofs.
Throughout, our notation and style of proof follow those in
Ref.~\cite{Cover-and-Thomas}.

\subsubsection{Entropy Defined}
\label{EntropyDefn}

Given a random variable $X$ taking values in a countable set ${\cal A}$, the
entropy of $X$ is
\begin{eqnarray}
\label{entropy-defined}
H[X] & \equiv & - \sum_{x \in {\cal A}}{\Prob(X = x) \log_2{\Prob(X = x)}} ~,
\end{eqnarray}
taking $0 \log{0} = 0$. Notice that $H[X]$ is the expectation value of
$- \log_2{\Prob(X = x)}$ and is measured in {\em bits} of information.
Caveats of the form ``when the sum converges to a finite value'' are implicit
in all statements about the entropies of infinite countable sets ${\cal A}$.

Shannon interpreted $H[X]$ as the {\it uncertainty in $X$}.  (Those leery of
any subjective component in notions like ``uncertainty'' may read ``effective
variability'' in its place.)  He showed, for example, that $H[X]$ is the mean
number of yes-or-no questions needed to pick out the value of $X$ on repeated
trials, if the questions are chosen to minimize this average \cite{Shannon-1948}.

\subsubsection{Joint and Conditional Entropies}
\label{JointCondEntropyDefn}

We define the joint entropy $H[X,Y]$ of two variables $X$ (taking values in
${\cal A}$) and $Y$ (taking values in ${\cal B}$) in the obvious way,
\begin{eqnarray}
& H & [X, Y] \equiv \\ \nonumber
	& - & \sum_{(x,y) \in {\cal A} \times {\cal B}}
	{\Prob(X = x \ProbAnd Y = y) \log_2{\Prob(X = x \ProbAnd Y = y)}} ~.
\label{joint-entropy-defined}
\end{eqnarray}
We define the conditional entropy $H[X|Y]$ of one random variable $X$ with
respect to another $Y$ from their joint entropy:
\begin{eqnarray}
H[X | Y] & \equiv & H[X, Y] - H[Y] ~.
\label{conditional-entropy-defined}
\end{eqnarray}
This also follows naturally from the definition of conditional probability,
since $\Prob(X = x | Y = y) \equiv \Prob(X = x \ProbAnd Y = y) / \Prob(Y = y)$.
$H[X|Y]$ measures the mean uncertainty remaining in $X$ once we know $Y$.

\subsubsection{Mutual Information}
\label{MutualInfoDefn}

The mutual information $I[X;Y]$ between two variables is defined to be
\begin{eqnarray}
I[X; Y] \equiv H[X] - H[X|Y] ~.
\label{def-of-mutual-info}
\end{eqnarray}
This is the average reduction in uncertainty about $X$ produced by fixing $Y$.
It is non-negative, like all entropies here, and symmetric in the two
variables.

\subsection{Patterns in Ensembles}
\label{PatternsInEnsembles}

It will be convenient to have a way of talking about the uncertainty of the
future.  Intuitively, this would just be $H[\Future]$, but in general that
quantity is infinite and awkward to manipulate.  (The special case in which
$H[\Future]$ is finite is dealt with in
App.~\ref{FiniteEntropyInfiniteFuture}.) Normally, we evade this by considering
$H[\FutureL]$, the uncertainty of the next $L$ symbols, treated as a function
of $L$.  On occasion, we will refer to the entropy per symbol or {\it entropy
rate} \cite{Shannon-1948,Cover-and-Thomas}:
\begin{eqnarray}
h[\Future] & \equiv & \lim_{\LLimit}{{1 \over L}H[\FutureL]} ~,
\label{EntropyRateBlockDefn}
\end{eqnarray}
and the {\em conditional entropy rate},
\begin{eqnarray}
h[\Future|X] & \equiv & \lim_{\LLimit}{{1 \over L}H[\FutureL|X]} ~,
\label{ConditionalEntropyRateDefn}
\end{eqnarray}
where $X$ is some random variable and the limits exist.  For stationary
stochastic processes, the limits always exist \cite[Theorem 4.2.1,
p.~64]{Cover-and-Thomas}.

These entropy rates are also always bounded above by $H[S]$; which is a special
case of Eq.~(\ref{IndepBound}).  Moreover, if $h[\Future] = H[S]$, the process
consists of independent variables---independent, identically distributed (IID)
variables, in fact, since we are only concerned with stationary processes here.

\begin{definition}[Capturing a Pattern]
\label{capturing-a-pattern}
$\AlternateStateSet$ {\it captures a pattern} if and only if there exists an
$L$ such that
\begin{equation}
H[\FutureL | \AlternateState] < LH[S] ~.
\end{equation}
\end{definition}
\addcontentsline{toc}{subsection}{\numberline{}Capturing a Pattern Defined}
This says that $\AlternateStateSet$ captures a pattern when it tells us
something about how the distinguishable parts of a process affect each other:
$\AlternateStateSet$ exhibits their dependence.  (We also speak of $\eta$,
the function associated with pasts, as capturing a pattern, since this is
implied by $\AlternateStateSet$ capturing a pattern.) Supposing that these
parts {\it do not} affect each other, then we have IID random variables, which
is as close to the intuitive notion of ``patternless'' as one is likely to
state mathematically.  Note that, because of the independence bound on joint
entropies (Eq.~(\ref{IndepBound})), if the inequality is satisfied for some
$L$, it is also satisfied for every ${L}^{\prime} > L$.  Thus, we can consider
the difference $H[S] - H[\FutureL | \AlternateState] / L$, for the smallest
$L$ for which it is nonzero, as the {\it strength of the pattern} captured by
$\AlternateStateSet$.  We will now mark an upper bound (Lemma
\ref{old-country-lemma}) on the strength of patterns; later we will show how to
attain this upper bound (Thm.~\ref{optimal-prediction-theorem}).

\subsection{The Lessons of History}

We are now in a position to prove a result about patterns in ensembles that
will be useful in connection with our later theorems about causal states.

\begin{lemma}[Old Country Lemma]
For all $\AlternateStateSet$ and for all $L \in {\Bbb Z}^{+}$,
\begin{equation}
H[\FutureL|\AlternateState] \geq H[\FutureL|\Past]  ~.
\label{cant-beat-the-past}
\end{equation}
\label{old-country-lemma}
\end{lemma}
\addcontentsline{toc}{subsection}{\numberline{}Old Country Lemma}

{\it Proof.}
By construction (Eq.~(\ref{EtaDefn})), for all $L$,
\begin{equation}
H[\FutureL|\AlternateState]  =  H[\FutureL|\eta(\Past)] ~.
\end{equation}
But
\begin{equation}
H[\FutureL|\eta(\Past)] \geq H[\FutureL|\Past] ~,
\end{equation}
since the entropy conditioned on a variable is never more than the entropy
conditioned on a function of the variable
(Eq.~(\ref{conditioning-on-function})).  QED.

{\it Remark 1.}
That is, conditioning on the whole of the past reduces the uncertainty in the
future to as small a value as possible.  Carrying around the whole semi-infinite
past is rather bulky and uncomfortable and is a somewhat dismaying prospect.
Put a bit differently: we want to forget as much of the past as possible and
so reduce its burden.  It is the contrast between this desire and the result of
Eq.~(\ref{cant-beat-the-past}) that leads us to call this the {\em Old Country
Lemma}.

{\it Remark 2.}
Lemma \ref{old-country-lemma} establishes the promised upper bound on
the strength of patterns: viz., the strength of the pattern is at most $H[S]
- H[\FutureL|\Past] / {L}_{past}$, where ${L}_{past}$ is the least value of $L$
such that $H[\FutureL|\Past] < LH[S]$.

\subsection[Minimality and Prediction]{Minimality and Prediction}

Let's invoke Occam's Razor: ``It is vain to do with more what can be done with
less'' \cite{Occam}.  To use the razor, we need to fix what is to be ``done''
and what ``more'' and ``less'' mean.  The job we want done is accurate
prediction, i.e., reducing the conditional entropies
$H[\FutureL|\AlternateState]$ as far as possible, the goal being
to attain the bound set by Lemma
\ref{old-country-lemma}.  But we want to do this as simply as possible, with as
few resources as possible.  On the road to meeting these two
constraints---minimal uncertainty and minimal resources---we will need a
measure of the second.  Since $\Prob(\Past = \past)$ is well defined, there is
an induced measure on the $\eta$-states; i.e.,
$\Prob(\AlternateState=\alternatestate)$, the probability of being in any
particular effective state, is well defined.  Accordingly, we define the
following measure of resources.

\begin{definition}
{\bf (Complexity of State Classes)}
The statistical complexity of a class $\AlternateStateSet$ of states is
\begin{eqnarray}
\Cmu(\AlternateStateSet) & \equiv & H[\AlternateState] \\ \nonumber
  & = & - \sum_{\alternatestate \in \AlternateState} \Prob(\AlternateState=\alternatestate) \log_2 \Prob(\AlternateState=\alternatestate) ~,
\end{eqnarray}
when the sum converges to a finite value.
\label{statistical-complexity-defined}
\end{definition}
\addcontentsline{toc}{subsection}{\numberline{}Complexity of State Classes}
The $\mu$ in $\Cmu$ reminds us that it is a measure-theoretic property and
depends ultimately on the distribution over the process's sequences, which
induces a measure over states.

The statistical complexity of a state class is the average uncertainty (in
bits) in the process's current state. This, in turn, is the same as the
average amount of memory (in bits) that the process {\em appears} to retain
about the past, given the chosen state class $\AlternateStateSet$. (We will
later, in Def.~\ref{statistical-complexity-of-a-process}, see how to define
the statistical complexity of a process itself.) The goal is to do with as
little of this memory as possible. Restated then, we want to minimize
statistical complexity, subject to the constraint of maximally accurate
prediction.

The idea behind calling the collection of all partitions of $\AllPasts$ Occam's
pool should now be clear: One wants to find the shallowest point in the pool.
This we now do.

\section{Computational Mechanics}

\begin{quotation}
Those who are good at archery learnt from the bow and not from Yi the Archer.
Those who know how to manage boats learnt from the boats and not from Wo.
%Those who can think learnt from themselves, and not from the Sages.
\\
\\
---Anonymous in Ref.~\cite{Kuan-Yin-Tzu}.
\end{quotation}

The ultimate goal of computational mechanics is to discern the patterns
intrinsic to a process.  That is, as much as possible, the goal is to let the
process describe itself, on its own terms, without appealing to {\em a priori}
assumptions about the process's structure.  Here we simply explore the
consistency and well-definedness of these goals.  Of course, practical
constraints may keep us from doing more than approximating these ideals more or
less grossly.  Naturally, such problems, which always turn up in implementation,
are much easier to address if we start from secure foundations.

\subsection{Causal States}
\label{DefnCausalStatesEMs}

\begin{definition}
{\bf (A Process's Causal States)}
The {\it causal states} of a process are the members of the range of the
function $\epsilon : \AllPasts \mapsto 2^\AllPasts$---the power set of
$\AllPasts$:
\begin{eqnarray}
\nonumber
\epsilon (\past) \equiv
  \{ \past^\prime | \Prob(\Future = \future & | & \Past = \past ) =
  \Prob(\Future = \future | \Past = \past^\prime )~, \\
  & & {\rm for~all} \future \in \Future , \past^\prime \in \Past\} ~,
\label{def-of-causal-states}
\end{eqnarray}
that maps from histories to sets of histories.  We write the $i^{th}$ causal
state as $\CausalState_i$ and the set of all causal states as
$\CausalStateSet$; the corresponding random variable is denoted $\CausalState$,
and its realization $\causalstate$.
\label{CausalStatesFunctionDefn}
\end{definition}
\addcontentsline{toc}{subsection}{\numberline{}Causal States of a Process Defined}

The cardinality of $\CausalStateSet$ is unspecified.  $\CausalStateSet$
can be finite, countably infinite, a continuum, a Cantor set, or something
stranger still.  Examples of these are given in Refs.~\cite{Calculi-of-emergence}
and \cite{Upper-thesis};
see especially the examples for hidden Markov models given there.

Alternately and equivalently, we could define an equivalence relation
$\CausalEquivalence$ such that two histories are equivalent if and only if they
have the same conditional distribution of futures, and then define causal
states as the equivalence classes generated by $\CausalEquivalence$.  (In fact,
this was the original approach \cite{Inferring-stat-compl}.)  Either way, the divisions of
this partition of $\AllPasts$ are made between regions that leave us in
different conditions of ignorance about the future.
\label{CausalStatesRelationDefn}

This last statement suggests another, still equivalent, description of
$\epsilon$:
\begin{eqnarray}
\nonumber
\epsilon(\past) = \{ \past^\prime
  | \Prob(\FutureL & = & \futureL | \Past = \past)
  = \Prob(\FutureL = \futureL | \Past = \past^\prime ) ~, \\
  \futureL \in \FutureL, && \past^\prime \in \Past ~, L \in {\Bbb Z}^{+} \} ~.
\label{finite-def-of-causal-states}
\end{eqnarray}
Using this we can make the original definition, Eq.
(\ref{def-of-causal-states}), more intuitive by picturing a sequence of
partitions of the space $\AllPasts$ of all histories in which
each new partition, induced using $L+1$,
is a refinement of the previous one induced using $L$.  At the coarsest
level, the first partition ($L=1$) groups together those histories that have
the same distribution for the very next observable.  These classes are then
subdivided using the distribution of the next two observables, then the next
three, four, and so on.  The limit of this sequence of partitions---the point at
which every member of each class has the same distribution of futures, of
whatever length, as every other member of that class---is the partition of
$\AllPasts$ induced by $\CausalEquivalence$.  See
App.~\ref{ReviewEquivalenceRelation} for a detailed discussion and review of
the equivalence relation $\CausalEquivalence$.

Although they will not be of direct concern in the following, due to the
time-asymptotic limits taken, there are transient causal states in addition to
those (recurrent) causal states defined above in
Eq.~(\ref{def-of-causal-states}).  Roughly speaking, the transient causal states
describe how a lengthening sequence (a history) of observations allows us to
identify the recurrent causal states with increasing precision.  See the
developments in App.~\ref{ReviewEquivalenceRelation} and in Refs.~\cite{Upper-thesis}
and \cite{DNCO} for more detail on transient causal states.

Causal states are a particular kind of effective state, and they have all the
properties common to effective states (Sec.~\ref{ThePool}).  In particular,
each causal state $\CausalState_i$ has several structures attached:
\begin{enumerate}
\item The index $i$---the state's ``name''.
\item The set of histories that have brought the process to
$\CausalState_i$, which we denote $\{ \past \in \CausalState_i \}$.
\item A conditional distribution over futures, denoted
$\Prob ( \Future | \CausalState_i)$, and equal to
$\Prob(\Future|\past), \past \in \CausalState_i$. Since we refer to this
type of distribution frequently and since it is the ``shape of the future'',
we call it the state's {\it morph}.
\end{enumerate}
Ideally, each of these should be denoted by a different symbol, and there
should be distinct functions linking each of these structures to their causal
state. To keep the growth of notation under control, however, we shall be
strategically vague about these distinctions. Readers may variously picture
$\epsilon$ as mapping histories to (i) simple indices, (ii) subsets of
histories, or (iii) ordered triples of indices, subsets, and morphs; or one
may even leave $\epsilon$ uninterpreted, as preferred, without interfering
with the development that follows.

\begin{figure}
\epsfxsize=2.7in
\begin{center}
\leavevmode
\epsffile{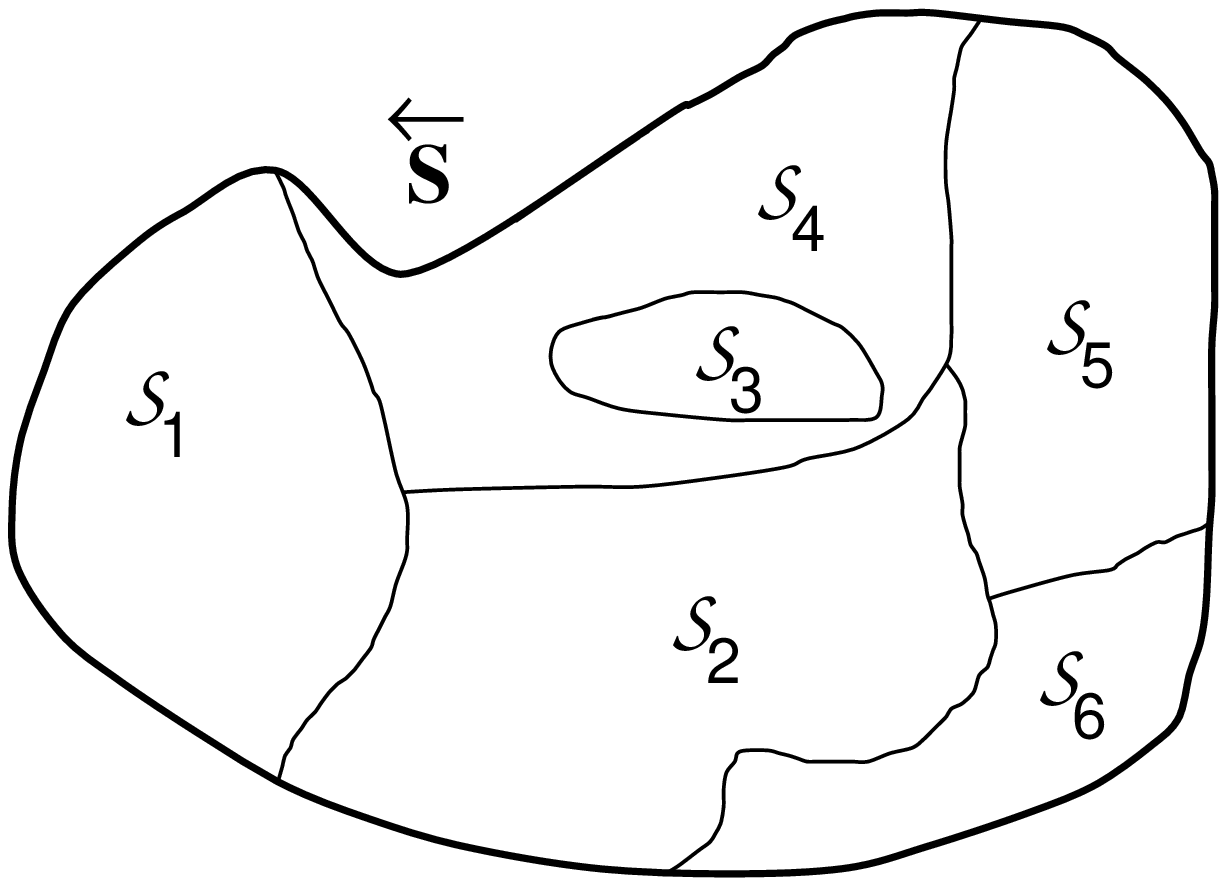}
\end{center}
\caption{A schematic representation of the partitioning of the set $\AllPasts$
of all histories into causal states $\CausalState_i \in \CausalStateSet$.
Within each causal state all the individual histories $\past$ have the same
morph---the same conditional distribution $\Prob(\Future | \past)$ for future
observables.
%  After Ref.~\cite{TDCS}.
  }
\label{epsilon-partition}
\end{figure}

\subsubsection{Morphs}

Each causal state has a unique morph, i.e., no two causal states have
the same conditional distribution of futures. This follows directly from
Def.~\ref{CausalStatesFunctionDefn}, and it is not true of effective states
in general. Another immediate consequence of that definition is that
\begin{equation}
\label{conditional-distributions-of-causal-states-and-histories-are-equal}
\Prob(\Future = \future | \CausalState = \epsilon(\past)) = \Prob(\Future = \future |
\Past = \past).
\end{equation}
(Again, this is not generally true of effective states.)  This observation lets
us prove a useful lemma about the conditional independence of the past
$\Past$ and the future $\Future$.

\begin{lemma}The past and the future are independent, conditioning on the
causal states.
\end{lemma}
\addcontentsline{toc}{subsubsection}{\numberline{}Independence of Past and
Future Conditional on a Causal State}

{\it Proof.}
Recall that two random variables $X$ and $Z$ are conditionally independent
if and only if there is a third variable $Y$ such that
\begin{eqnarray}
\nonumber
\Prob( X & = & x \ProbAnd Y = y \ProbAnd Z = z) \\
         & = & \Prob(X = x | Y = y) \Prob(Z = z | Y = y) \Prob(Y = y) ~.
\end{eqnarray}
That is, all of the dependence of $Z$ on $X$ is mediated by $Y$.  For
convenience below we note that, re-factoring the conditional probabilities,
this is equivalent to the requirement that:
\begin{eqnarray}
\nonumber
\Prob( X & = & x \ProbAnd Y = y \ProbAnd Z = z) \\
  & = & \Prob(Z = z | Y = y) \Prob(Y = y | X = x) \Prob(X = x) ~.
\end{eqnarray}

Let us consider
$\Prob(\Past = \past \ProbAnd \CausalState = \causalstate \ProbAnd \Future = \future)$.
\begin{eqnarray}
\nonumber
\Prob & ( & \Past = \past \ProbAnd \CausalState = \causalstate \ProbAnd \Future = \future) \\
& = & \Prob(\Future = \future | \CausalState = \causalstate \ProbAnd \Past = \past)
	\Prob(\CausalState = \causalstate \ProbAnd \Past = \past)
\label{partial-decomposition}\\ \nonumber
& = & \Prob(\Future = \future | \CausalState = \causalstate \ProbAnd \Past = \past)
	\Prob(\CausalState = \causalstate | \Past = \past) \Prob(\Past = \past) ~.
\end{eqnarray}
Now, $\Prob(\CausalState = \causalstate | \Past = \past) = 0$, unless $\causalstate = \epsilon(\past)$, 
which case $\Prob(\CausalState = \causalstate | \Past = \past) = 1$.  Either way, the first two factors in the last line of Eq.~(\ref{partial-decomposition}) can be
written, by
Eq.~(\ref{conditional-distributions-of-causal-states-and-histories-are-equal}),
\begin{eqnarray}
\nonumber
\Prob & ( & \Future = \future | \CausalState = \causalstate \ProbAnd \Past = \past)
  \Prob(\CausalState = \causalstate | \Past = \past) \\
  & = & \Prob(\Future = \future | \CausalState = \causalstate) \Prob(\CausalState = \causalstate | \Past = \past) ~,
\label{advanced-decomposition}
\end{eqnarray}
so that, substituting Eq.~(\ref{advanced-decomposition}) into
Eq.~(\ref{partial-decomposition}),
\begin{eqnarray}
\nonumber
\Prob & ( & \Past = \past \ProbAnd \CausalState = \causalstate \ProbAnd \Future = \future) \\
  & = & \Prob(\Future = \future | \CausalState = \causalstate)
	\Prob(\CausalState = \causalstate | \Past = \past)\Prob(\Past = \past) ~.
\end{eqnarray}
QED.

\subsubsection{Homogeneity}

Following Ref.~\cite{Salmon-1984}, we introduce two new definitions and a lemma
which are required later on, especially in the proof of Lemma
\ref{refinement-lemma} and the theorems depending on that lemma.

\begin{definition}
{\bf (Strict Homogeneity)}
A set $\bf X$ is {\em strictly homogeneous} with respect to a certain random
variable $Y$ when the conditional distribution $\Prob(Y|{\bf X})$ for $Y$ is
the same for all subsets of $\bf X$.
\label{StrictHomogeneityDefn}
\end{definition}
\addcontentsline{toc}{subsubsection}{\numberline{}Strict Homogeneity}

\begin{definition}
{\bf (Weak Homogeneity)}
A set $\bf X$ is {\em weakly homogeneous} with respect to $Y$ if $\bf X$ is not
strictly homogeneous with respect to $Y$, but
${\bf X} \setminus {\bf X}_0$ ($\bf X$ with ${\bf X}_0$ removed) is, where
${\bf X}_0$ is a subset of $\bf X$ of measure $0$.
\label{WeakHomogeneityDefn}
\end{definition}
\addcontentsline{toc}{subsubsection}{\numberline{}Weak Homogeneity}

\begin{lemma}
{\bf (Strict Homogeneity of Causal States)}
A process's causal states are the largest subsets of histories that are
all strictly homogeneous with respect to futures of all lengths.
\label{homogeneity-lemma}
\end{lemma}
\addcontentsline{toc}{subsubsection}{\numberline{}Strict Homogeneity of Causal
States}

{\it Proof.} We must show that, first, the causal states are strictly
homogeneous with respect to futures of all lengths and, second, that no
larger strictly homogeneous subsets of histories could be made. The first
point, the strict homogeneity of the causal states, is evident from
Eq.~(\ref{finite-def-of-causal-states}): By construction, all elements of
a causal state have the same morph, so any part of a causal state will
have the same morph as the whole state. The second point likewise follows
from Eq.~(\ref{finite-def-of-causal-states}), since the causal state by
construction contains {\em all} the histories with a given morph. Any other
set strictly homogeneous with respect to futures must be smaller than a causal
state, and any set that includes a causal state as a proper subset cannot be
{\em strictly} homogeneous. QED.

{\it Remark.}  The statistical explanation literature would say that causal
states are the ``statistical-relevance basis for causal explanations''.  The
elements of such a basis are, precisely, the largest classes of combinations of
independent variables with homogeneous distributions for the dependent
variables.  See Ref.~\cite{Salmon-1984} for further discussion along these lines.

\subsection{Causal State-to-State Transitions}

The causal state at any given time and the next value of the observed
process together determine a new causal state; this is proved shortly in Lemma
\ref{automatism-lemma}.  Thus, there is a natural relation of succession among
the causal states; recall the discussion of causality in Sec.~\ref{Causation}.
Moreover, given the current causal state, all the possible next
values have well defined conditional probabilities.  In fact, by construction
the entire semi-infinite future does.  Thus, there is a well defined probability
${T}_{ij}^{(s)}$ of the process generating the value $s \in {\cal A}$ and going
to causal state $\CausalState_j$, if it is in state $\CausalState_i$.

\begin{definition}
{\bf (Causal Transitions)}
The labeled transition probability $T_{ij}^{(s)}$ is the probability of
making the transition from state $\CausalState_i$ to state $\CausalState_j$ while emitting
the symbol $s \in {\cal A}$:
\begin{equation}
T_{ij}^{(s)} \equiv
  \Prob(\CausalState^\prime = \CausalState_j \ProbAnd \Future^1 = s | \CausalState = \CausalState_i ) ~,
\end{equation}
\label{CausalTransitionsDefn}
\end{definition}
\addcontentsline{toc}{subsection}{\numberline{}Causal Transitions}
where $\CausalState$ is the current causal state and $\CausalState^\prime$
its successor on emitting $s$. We denote the set
$\{ T_{ij}^{(s)} : s \in {\cal A} \}$ by $\bf T$.

\begin{lemma}[Transition Probabilities]
$T_{ij}^{(s)}$ is given by
\begin{eqnarray}
T_{ij}^{(s)} & = & \Prob(\past s \in {\CausalState}_{j} | \past \in {\CausalState}_{i}) \\
	& = & {\Prob(\past \in {\CausalState}_{i} \ProbAnd \past s \in {\CausalState}_{j}) \over
	\Prob(\past \in {\CausalState}_{i})} ~,
\end{eqnarray}
where $\past s$ is read as the semi-infinite sequence obtained by concatenating
$s \in {\cal A}$ onto the end of $\past$.
\end{lemma}
\addcontentsline{toc}{subsection}{\numberline{}Transition Probabilities}

{\it Proof.}
\begin{eqnarray}
T^{(s)}_{ij} & = & \Prob(\CausalState^\prime = \CausalState_j
   \ProbAnd \Future^1 = s | \CausalState = \CausalState_i) \\
  & = & { \Prob(\CausalState^\prime = \CausalState_j
  \ProbAnd \Future^1 = s \ProbAnd \CausalState = \CausalState_i) }
  \over { \Prob(\CausalState = \CausalState_i) } ~.
\label{TransProbIntermediate}
\end{eqnarray}
Now $\CausalState = \CausalState_i$ if and only if $\past \in \CausalState_i$,
and $\CausalState^\prime = \CausalState_j$ if and only
$\past^\prime \in \CausalState_j$, where by
$\past^\prime$ we mean the history that is the immediate successor to $\past$;
for consistency, $\past^\prime = \past s$.
So we can rewrite Eq.~(\ref{TransProbIntermediate}) as
\begin{eqnarray}
T^{(s)}_{ij} & = & { \Prob(\past \in \CausalState_i
  \ProbAnd \Future^1 = s \ProbAnd \past^\prime \in \CausalState_j) }
  \over { \Prob(\CausalState = \CausalState_i) } \\
  & = & { \Prob(\past \in \CausalState_i \ProbAnd \Future^1 = s
  \ProbAnd \past s \in \CausalState_j) } \over { \Prob(\CausalState = \CausalState_i) } \\
  & = & { \Prob(\past \in \CausalState_i \ProbAnd \past s \in \CausalState_j) }
  \over { \Prob(\CausalState = \CausalState_i) } ~.
\end{eqnarray}
In the third line we used the fact that $\Past = \past$ and $\Past^\prime =
\past s$ jointly imply $\Future^1 = s$, making that condition redundant.  QED.

Notice that $T_{ij}^{(\lambda)} = \delta_{ij}$; that is, the transition labeled
by the null symbol $\lambda$ is the identity.

\subsection{$\epsilon$-Machines}

The combination of the function $\epsilon$ from histories to causal states
with the labeled transition probabilities $T_{ij}^{(s)}$ is called the
{\it $\epsilon$-machine} of the process \cite{Calculi-of-emergence,Inferring-stat-compl}.

\begin{definition}[An $\epsilon$-Machine Defined]
The $\epsilon$-machine of a process is the ordered pair
$\{ \epsilon, {\bf T} \}$, where $\epsilon$ is the causal state function and
$\bf T$ is set of the transition matrices for the states defined by $\epsilon$.
\end{definition}
\addcontentsline{toc}{subsection}{\numberline{}An $\epsilon$-Machine Defined}

Equivalently, we may denote an $\epsilon$-machine by
$\{ \CausalStateSet, {\bf T} \}$.

To satisfy the algebraic requirement outlined in Sec.~\ref{Desiderata}, we make
explicit the connection with semi-group theory.

\begin{proposition}[$\epsilon$-Machines Are Monoids]
The algebra generated by the $\epsilon$-machine $\{ \epsilon, {\bf T} \}$
is a semi-group with an identity element, i.e., it is a monoid.
\label{eMSemiGroup}
\end{proposition}
\addcontentsline{toc}{subsection}{\numberline{}$\epsilon$-Machines Are Monoids}

{\it Proof.}
See App.~\ref{SemiGroups}.

{\it Remark.} Due to this, $\epsilon$-machines can be interpreted as
capturing a process's {\em generalized symmetries}.  Any subgroups of an
$\epsilon$-machine's semi-group are, in fact, symmetries in the more
familiar sense.

\begin{lemma}
{\bf ($\epsilon$-Machines Are Deterministic)}
For each $\CausalState_i$ and $s \in {\cal A}$, ${T}^{(s)}_{ij} > 0$ only
for that $\CausalState_j$ for which $\epsilon(\past \! s) = \CausalState_j$
if and only if $\epsilon(\past) = \CausalState_i$, for all pasts $\past$.
\label{automatism-lemma}
\end{lemma}
\addcontentsline{toc}{subsection}{\numberline{}$\epsilon$-Machines Are
Deterministic}

{\it Proof.}
The lemma is equivalent to asserting that for all $s \in {\cal A}$ and
$\past, {\past}^{\prime} \in \AllPasts$, if
$\epsilon(\past) = \epsilon({\past}^{\prime})$, then
$\epsilon(\past \! s) = \epsilon({\past}^{\prime} \! s)$.
($\past \! s$ is just another history and belongs to one or another
causal state.)

Suppose this were not true.  Then there would have to exist at least one future
$\future$ such that
\begin{eqnarray}
\Prob(\Future = \future | \Past = \past \! s ) & \neq &
  \Prob(\Future = \future | \Past = \past^\prime \! s ) ~,
\end{eqnarray}
when nonetheless $\epsilon(\past) = \epsilon(\past^\prime)$.  Equivalently, we
would have
\begin{equation}
{\Prob(\BiInfinity = \past \! s \! \future) \over \Prob(\Past = \past \! s)}
  \neq {\Prob(\BiInfinity = \past^\prime \! s \! \future)
  \over \Prob(\Past = \past^\prime \! s)} ~,
\end{equation}
where we read $s \! \future$ as the semi-infinite string that begins $s$ and
continues $\future$.  (Remember, the point at which we break the
stochastic process into a past and a future is arbitrary.) However, the
probabilities in the denominators are equal to $\Prob(\Future^1 = s|\Past =
\past) \Prob(\Past = \past)$ and $\Prob(\Future^1 = s|\Past =
\past^\prime)\Prob(\Past = \past^\prime)$, respectively, and by assumption
$\Prob(\Future^1 = s|\Past = \past^\prime)=\Prob(\Future^1 = s|\Past = \past)$,
since $\epsilon(\past^\prime) = \epsilon(\past)$.  Therefore, we would need
\begin{equation}
{\Prob(\BiInfinity = \past s \future) \over \Prob(\Past = \past)}
  \neq {\Prob(\BiInfinity = \past^\prime \! s \! \future)
  \over \Prob(\Past = \past^\prime)} ~.
\end{equation}
This is the same, though, as
\begin{equation}
\Prob(\Future = s \! \future| \Past = \past)
  \neq \Prob(\Future =  s \! \future | \Past = \past^\prime) ~.
\end{equation}
This is to say that there is a future $s \! \future$ that has different
probabilities depending on whether we conditioned on $\past$ or on
${\past}^{\prime}$.  But this contradicts the assumption that the two histories
belong to the same causal state.  Therefore, there is no such future $\future$,
and the alternative statement of the lemma is true.  QED.

{\it Remark 1.}
In automata theory \cite{Hopcroft-Ullman}, a set of states and transitions is said to be
{\it deterministic} if the current state and the next input---here, the next
result from the original stochastic process---together fix the next state.
This use of the word ``deterministic'' is often confusing, since many
stochastic processes (e.g., simple Markov chains) are deterministic in this
sense.

{\it Remark 2.}
Starting from a fixed state, a given symbol always leads to at most one single
state.  But there can be several transitions from one state to another, each
labeled with a different symbol.

{\it Remark 3.}
Clearly, if $T^{(s)}_{ij} > 0$, then
$T^{(s)}_{ij} = \Prob(\Future^1 = s | \CausalState = \CausalState_i)$.
In automata theory the ``disallowed'' transitions
($T^{(s)}_{ij} = 0$) are sometimes explicitly represented and lead to a
``reject'' state indicating that the particular history does not occur.

\begin{lemma}
{\bf (Causal States Are Independent)}
The probability distributions over causal states at different times are
conditionally independent.
\label{independent-states-lemma}
\end{lemma}
\addcontentsline{toc}{subsection}{\numberline{}Causal States Are Independent}

{\it Proof.}
What we wish to show is that, writing $\CausalState$, $\CausalState^\prime$,
$\CausalState^{\prime\prime}$ for the sequence of causal states at three
successive times, $\CausalState$ and ${\CausalState}^{\prime\prime}$ are
conditionally independent, given ${\CausalState}^{\prime}$.  We can do this
directly:
\begin{eqnarray}
\nonumber
\Prob & ( & \CausalState = \causalstate \ProbAnd \CausalState^\prime = \causalstate^\prime \ProbAnd
  \CausalState^{\prime\prime} = \causalstate^{\prime\prime}) \\
\nonumber
& = & \Prob(\CausalState^{\prime\prime} = \causalstate^{\prime\prime} | \CausalState = \causalstate
  \ProbAnd \CausalState^\prime = \causalstate^\prime)
  \Prob(\CausalState = \causalstate \ProbAnd \CausalState^\prime = \causalstate^\prime) \\
& = & \Prob(\Future^1 \in a | \CausalState = \causalstate \ProbAnd \CausalState^\prime = \causalstate^\prime)
  \Prob(\CausalState = \causalstate \ProbAnd \CausalState^\prime = \causalstate^\prime) ~,
\end{eqnarray}
where $a$ is the subset of all symbols that lead from
$\causalstate^\prime$ to $\causalstate^{\prime\prime}$.  This is a well defined
subset, in virtue of Lemma \ref{automatism-lemma} immediately preceding, which
also guarantees the equality of conditional probabilities we have
used.  Likewise,
\begin{equation}
\Prob(\CausalState^{\prime\prime} = \causalstate^{\prime\prime} | \CausalState^\prime = \causalstate^\prime)
	= \Prob(\Future^1 \in a | \CausalState^\prime = \causalstate^\prime) ~.
\end{equation}
But, by construction,
\begin{equation}
\Prob(\Future^1 \in a | \CausalState = \causalstate \ProbAnd \CausalState^\prime = \causalstate^\prime)
	= \Prob(\Future^1 \in a | \CausalState^\prime = \causalstate^\prime) ~,
\end{equation}
and hence
\begin{equation}
\Prob(\CausalState^{\prime\prime} = \causalstate^{\prime\prime} | \CausalState^\prime = \causalstate^\prime)
	= \Prob({\CausalState}^{\prime\prime} = \causalstate^{\prime\prime}
	| \CausalState = \causalstate \ProbAnd \CausalState^\prime = \causalstate^\prime) ~.
\end{equation}
So, to resume,
\begin{eqnarray}
\nonumber
\Prob & ( & \CausalState = \causalstate \ProbAnd \CausalState^\prime = \causalstate^\prime \ProbAnd
  \CausalState^{\prime\prime} = \causalstate^{\prime\prime}) \\ \nonumber
& = & \Prob(\CausalState^{\prime\prime} = \causalstate^{\prime\prime}
  | \CausalState^\prime = \causalstate^\prime)
  \Prob(\CausalState = \causalstate \ProbAnd \CausalState^\prime = \causalstate^\prime) \\
& = & \Prob(\CausalState^{\prime\prime} = \causalstate^{\prime\prime} | \CausalState^\prime
  = \causalstate^\prime) \Prob(\CausalState^\prime = \causalstate^\prime | \CausalState = \causalstate)
  \Prob(\CausalState = \causalstate) ~.
\end{eqnarray}
The last line follows from the definition of conditional probability
and is equivalent to the more easily interpreted expression given by
\begin{equation}
\Prob( \CausalState^{\prime\prime} | \CausalState^\prime )
\Prob( \CausalState | \CausalState^\prime ) \Prob( \CausalState^\prime ) ~.
\label{CausalStateCondIndep}
\end{equation}
Thus, applying mathematical induction to Eq. (\ref{CausalStateCondIndep}),
causal states at different times are independent, conditioning on the
intermediate causal states. QED.

{\it Remark 1.}
This lemma strengthens the claim that the causal states are, in fact, the
causally efficacious states: given knowledge of the present state, what has
gone before makes no difference.  (Again, recall the philosophical
preliminaries of Sec.~\ref{Causation}.)

{\it Remark 2.}
This result indicates that the causal states, considered as a process, define a
kind of Markov chain.  Thus, causal states can be roughly considered to be a
generalization of Markovian states.  We say ``kind of'' since the class of
$\epsilon$-machines is substantially richer \cite{Calculi-of-emergence,Upper-thesis} than what
one normally associates with Markov chains \cite{Kemeny-finite-chains,Kemeny-denumerable-chains}.

\begin{definition}[$\epsilon$-Machine Reconstruction]
$\epsilon$-{\rm Machine reconstruction} is any procedure that given a process
$\Prob(\BiInfinity)$, or an approximation of $\Prob(\BiInfinity)$, produces
the process's $\epsilon$-machine $\{\CausalStateSet, {\bf T} \}$.
\label{ReconstructionDefn}
\end{definition}
\addcontentsline{toc}{subsection}{\numberline{}$\epsilon$-Machine Reconstruction}

Given a mathematical description of a process, one can often calculate
analytically its $\epsilon$-machine. (For example, see the computational
mechanics analysis of spin systems in Ref. \cite{DNCO}.)
There is also a wide range of algorithms
which reconstruct $\epsilon$-machines from empirical estimates of
$\Prob(\BiInfinity)$.  Some, such as those used in Refs.~\cite{Calculi-of-emergence,Inferring-stat-compl,Computation-at-the-onset,Hanson-thesis}, operate in ``batch'' mode, taking the
raw data as a whole and producing the $\epsilon$-machine.  Others could operate
incrementally, in ``on-line'' mode, taking in individual measurements and
re-estimating the set of causal states and their transition probabilities.

\section{Optimalities and Uniqueness}

We now show that: causal states are maximally accurate predictors of minimal
statistical complexity; they are unique in sharing both properties; and
their state-to-state transitions are minimally stochastic.  In other words,
they satisfy both of the constraints borrowed from Occam, and they are the only
representations that do so.  The overarching moral here is that causal states
and $\epsilon$-machines are {\em the} goals in any learning or modeling
scheme.  The argument is made by the time-honored means of proving optimality
theorems.  We address, in our concluding remarks (Sec.~\ref{Conclusion}), the
practicalities involved in attaining these goals.

\begin{figure}
\epsfxsize=2.7in
\begin{center}
\leavevmode
\epsffile{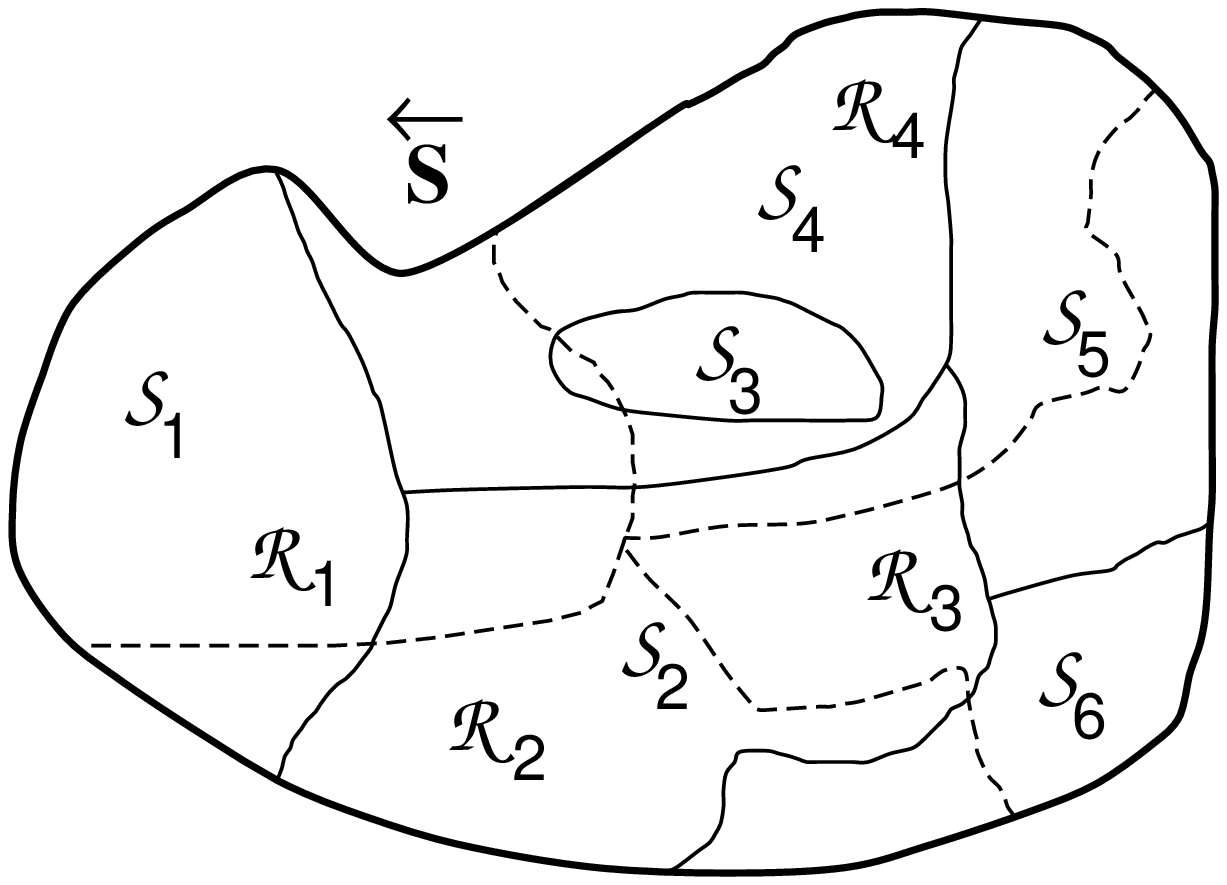}
\end{center}
\caption{An alternative class $\AlternateStateSet$ of states (delineated by
dashed lines) that partition $\AllPasts$ overlaid on the causal states
$\CausalStateSet$ (outlined by solid lines).  Here, for example,
$\CausalState_2$ contains parts of $\AlternateState_1$, $\AlternateState_2$,
$\AlternateState_3$ and $\AlternateState_4$.  The collection of all such
alternative partitions form {\em Occam's pool}.  Note again that the
$\AlternateState_i$ need not be compact nor simply connected, as drawn.
%  After Ref.~\cite{TDCS}.
  }
\label{epsilon-and-bad-eta-partition}
\end{figure}

As part of our strategy, though, we also prove several results that are not
optimality results; we call these lemmas to indicate their subordinate
status.  All of our theorems, and some of our lemmas, will be established by
comparing causal states, generated by $\epsilon$, with other rival sets of
states, generated by other functions $\eta$.  In short, none of the rival
states---none of the other patterns---can out-perform the causal states.

It is convenient to fix some additional notation.  Let $\CausalState$ be the
random variable for the current causal state, $\Future^1 \in {\cal A}$ the next
``observable'' we get from the original stochastic process,
$\CausalState^\prime$ the next causal state, $\AlternateState$ the current
state according to $\eta$, and $\AlternateState^\prime$ the next
$\eta$-state.  $\causalstate$ will stand for a particular value (causal state)
of $\CausalState$ and $\alternatestate$ a particular value of
$\AlternateState$.  When we quantify over alternatives to the causal states, we
quantify over $\AlternateStateSet$.

%\begin{theorem}[Causal States Are Maximally Prescient]
\begin{theorem}
{\bf (Causal States are Maximally Prescient)}
\cite{TDCS}

For all $\AlternateStateSet$ and all $L \in {\Bbb Z}^{+}$,
\begin{equation}
H[\FutureL|\AlternateState] \geq H[\FutureL|\CausalState] ~.
\end{equation}
\label{optimal-prediction-theorem}
\end{theorem}
\addcontentsline{toc}{subsection}{\numberline{}Causal States Are Maximally
Prescient}

{\it Proof.}
We have already seen that $H[\FutureL|\AlternateState] \geq H[\FutureL|\Past]$ (Lemma \ref{old-country-lemma}).  But by construction
(Def.~\ref{CausalStatesFunctionDefn}),
\begin{equation}
\Prob(\FutureL = \futureL|\Past = \past) =
  \Prob(\FutureL = \futureL | \CausalState = \epsilon(\past)) ~.
\end{equation}
Since entropies depend only on the probability distribution, $H[\FutureL|\CausalState]
= H[\FutureL|\Past]$ for every $L$.  Thus, $H[\FutureL|\AlternateState] \geq
H[\FutureL|\CausalState]$, for all $L$.  QED.

{\it Remark.}
That is to say, causal states are as good at predicting the future---are as
{\it prescient}---as complete histories.  In this, they satisfy the first
requirement borrowed from Occam.  Since the causal states are well defined and
since they can be systematically approximated, we have shown that the upper
bound on the strength of patterns (Def.~\ref{capturing-a-pattern} and Lemma
\ref{old-country-lemma}, Remark) can in fact be reached.  Intuitively, the
causal states achieve this because, unlike effective states in general, they do
not throw away any information about the future which might be contained in
$\Past$.  Even more colloquially, to paraphrase the definition of information
in Ref.~\cite{Bateson-mind-and-nature}, the causal states record every difference (about the
past) that makes a difference (to the future).  We can actually make this
intuition quite precise, in an easy corollary to the theorem.

\begin{corollary}
{\bf (Causal States Are Sufficient Statistics)}
The causal states $\CausalStateSet$ of a process are sufficient statistics for
predicting it.
\end{corollary}
\addcontentsline{toc}{subsection}{\numberline{}Causal States Are Sufficient Statistics}

{\it Proof.}
It follows from Thm.~\ref{optimal-prediction-theorem} and Eq.~(\ref{def-of-mutual-info}) that, for all $L \in {\Bbb Z}^{+}$,
\begin{equation}
I[\FutureL;\CausalState] = I[\FutureL;\Past] ~,
\end{equation}
where $I$ was defined in Eq.~(\ref{def-of-mutual-info}).  Consequently, the
causal state is a {\it sufficient statistic}---see Refs.~\cite[p.~37]{Cover-and-Thomas}
and \cite[sec.~2.4--2.5]{Kullback-info-theory-and-stats}---for predicting futures of any length.  QED.

All subsequent results concern rival states that are as prescient as the
causal states.  We call these {\em prescient rivals} and denote a class of
them $\PrescientStateSet$.

\begin{definition}[Prescient Rivals]
{\em Prescient rivals} $\PrescientStateSet$ are states that are as
predictive as the causal states; viz., for all $L \in {\Bbb Z}^{+}$,
\begin{equation}
H[\FutureL|\PrescientState] = H[\FutureL|\CausalState] ~.
\end{equation}
\label{PrescientRivals}
\end{definition}
\addcontentsline{toc}{subsection}{\numberline{}Prescient Rivals Defined}

{\it Remark.}
Prescient rivals are also sufficient statistics.

\begin{lemma}[Refinement Lemma]
For all prescient rivals $\PrescientStateSet$ and for each
$\prescientstate \in \PrescientStateSet$, there is a
$\causalstate \in \CausalStateSet$ and a measure-$0$
subset $\prescientstate_0 \subset \prescientstate$, possibly empty,
such that $\prescientstate \setminus \prescientstate_{0} \subseteq \causalstate$,
where $\setminus$ is set subtraction.
\label{refinement-lemma}
\end{lemma}
\addcontentsline{toc}{subsection}{\numberline{}Refinement Lemma}

{\it Proof.}
We invoke a straightforward extension of Thm.~2.7.3 of Ref.~\cite{Cover-and-Thomas}: If
$X_1, X_2, \ldots, X_n$ are random variables over the same set ${\cal A}$, each
with distinct probability distributions, $\Theta$ a random variable over the
integers from $1$ to $n$ such that $\Prob(\Theta = i) = \lambda_i$, and $Z$ a
random variable over ${\cal A}$ such that $Z = X_\Theta$, then
\begin{eqnarray}
\nonumber
H[Z] & = & H[\sum_{i=1}^n \lambda_i X_i ] \\
    & \geq & \sum_{i=1}^n \lambda_i H[X_i]  ~.
\label{EntropyConvexity}
\end{eqnarray}
In words, the entropy of a mixture of distributions is at least the mean of the
entropies of those distributions. This follows since $H$ is strictly concave,
which in turn follows from $x \log{x}$ being strictly convex for $x \geq 0$.
We obtain equality in Eq.~(\ref{EntropyConvexity}) if and only if all the
$\lambda_i$ are either 0 or 1, i.e., if and only if $Z$ is at least weakly
homogeneous (Def.~\ref{WeakHomogeneityDefn}).

The conditional distribution of futures for each rival state $\alternatestate$
can be written as a weighted mixture of the morphs of one or more causal
states.  (Cf.~Fig.~\ref{epsilon-and-bad-eta-partition}.)  Thus, by
Eq.~(\ref{EntropyConvexity}), unless every $\alternatestate$ is at least weakly
homogeneous with respect to $\FutureL$ (for each $L$), the entropy of
$\FutureL$ conditioned on $\AlternateState$ will be higher than the minimum,
the entropy conditioned on $\CausalState$.  So, in the case of the maximally
predictive $\PrescientState$, every $\prescientstate \in \PrescientStateSet$
must be at least weakly homogeneous with respect to all $\FutureL$.  But the
causal states are the largest classes that are strictly homogeneous with
respect to all $\FutureL$ (Lemma \ref{homogeneity-lemma}).  Thus, the strictly
homogeneous part of each $\prescientstate \in \PrescientStateSet$ must be a
subclass, possibly improper, of some causal state $\causalstate \in
\CausalStateSet$.  QED.

{\it Remark 1.}
An alternative proof appears in App.~\ref{AltProof}.

\begin{figure}
\epsfxsize=2.7in
\begin{center}
\leavevmode
\epsffile{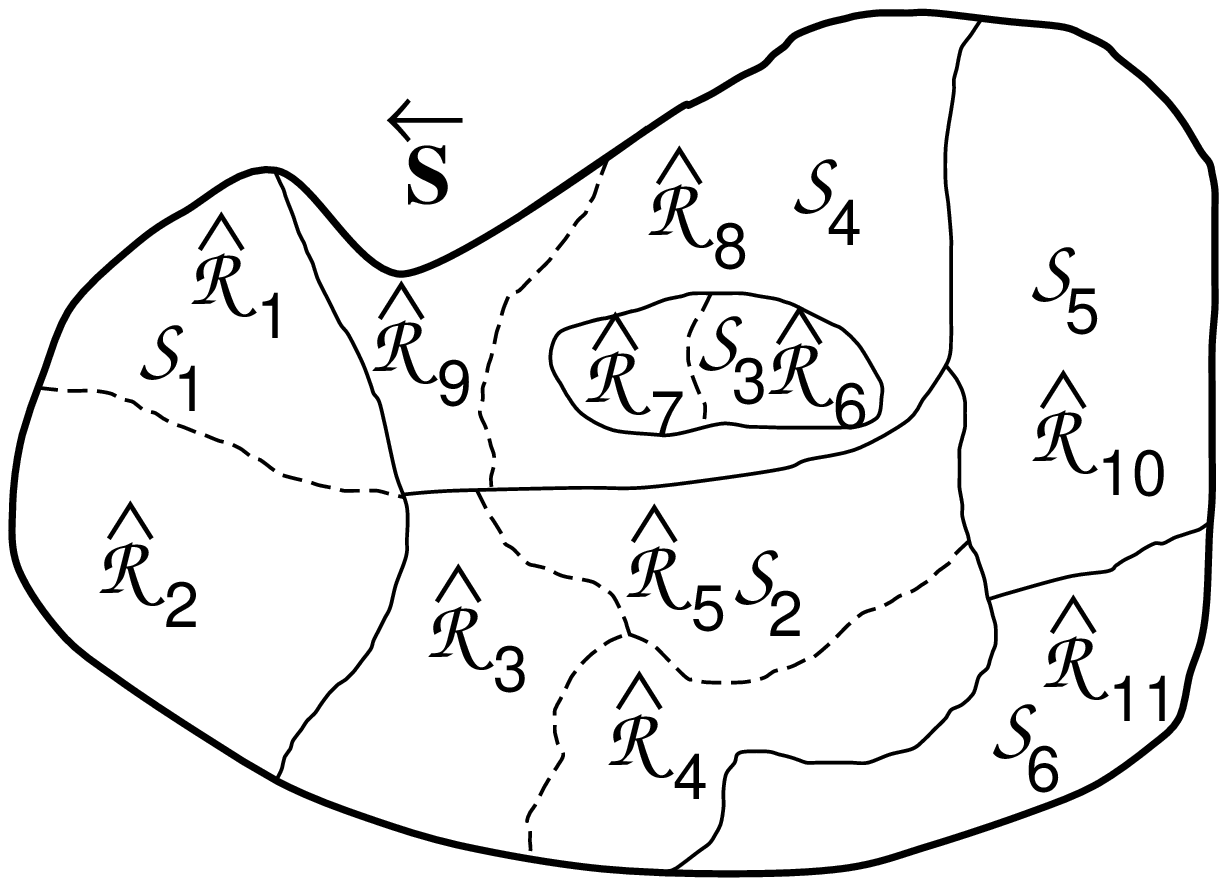}
\end{center}
\caption{A prescient rival partition $\PrescientStateSet$ must be a refinement
of the causal-state partition {\em almost everywhere}.  That is, almost all of
each $\PrescientState_i$ must contained within some $\CausalState_j$; the
exceptions, if any, are a set of histories of measure $0$.  Here for instance
$\CausalState_2$ contains the positive-measure parts of $\PrescientState_3$,
$\PrescientState_4$, and $\PrescientState_5$.  One of these rival states, say
$\PrescientState_3$, could have member-histories in any or all of the other
causal states, provided the total measure of such exceptional histories is zero.
Cf.~Fig.~\ref{epsilon-and-bad-eta-partition}.}
\label{refined-partition}
\end{figure}

{\it Remark 2.} 
The content of the lemma can be made quite intuitive, if we ignore for a moment
the measure-$0$ set $\prescientstate_0$ of histories mentioned in its statement.
It then asserts
that any alternative partition $\PrescientStateSet$ that is as prescient as the
causal states must be a refinement of the causal-state partition.  That is,
each $\PrescientState_i$ must be a (possibly improper) subset of some
$\CausalState_j$.  Otherwise, at least one $\PrescientState_i$ would have to
contain parts of at least two causal states.  And so, using this
$\PrescientState_i$ to predict the future observables would lead to more
uncertainty about $\Future$ than using the causal states.  This is illustrated
by Fig.~\ref{refined-partition}, which should be contrasted with
Fig.~\ref{epsilon-and-bad-eta-partition}.

Adding the measure-$0$ set $\prescientstate_0$ of histories to this picture
does not change its heuristic content much. Precisely because these histories
have zero probability, treating them in an ``inappropriate'' way makes no
discernible difference to predictions, morphs, and so on. There is a problem
of terminology, however, since there seems to be no standard name for the
relationship between the partitions $\PrescientStateSet$ and $\CausalStateSet$.
We propose to say that the former is a refinement of the latter
{\em almost everywhere} or, simply, a {\em refinement a.e.}

{\it Remark 3.}
One cannot work the proof the other way around to show that the causal states
have to be a refinement of the equally prescient $\PrescientState$-states.
This is precluded because applying the theorem borrowed from
Ref.~\cite{Cover-and-Thomas}, Eq.~(\ref{EntropyConvexity}), hinges on being able to
reduce uncertainty by specifying from {\it which} distribution one chooses.
Since the causal states are constructed so as to be strictly homogeneous with
respect to futures, this is not the case. Lemma~\ref{homogeneity-lemma}
and Thm.~\ref{optimal-prediction-theorem} together protect us.

{\it Remark 4.}
Because almost all of each prescient rival state is wholly contained within a
single causal state, we can construct a function $g: \PrescientStateSet \mapsto
\CausalStateSet$, such that, if $\eta(\past) = \prescientstate$, then
$\epsilon(\past) = g(\prescientstate)$ almost always.  We can even say that
$\CausalState = g(\PrescientState)$ almost always, with the understanding that
this means that, for each $\prescientstate$, $\Prob(\CausalState =
\causalstate| \PrescientState = \prescientstate) > 0$ if and only if
$\causalstate = g(\prescientstate)$.

\vspace{5mm}

\begin{theorem}[Causal States Are Minimal]
\cite{TDCS} For all prescient rivals $\PrescientStateSet$,
\begin{equation}
\Cmu(\PrescientStateSet) \geq \Cmu(\CausalStateSet) ~.
\end{equation}
\label{minimality-theorem}
\end{theorem}
\addcontentsline{toc}{subsection}{\numberline{}Causal States Are Minimal}

{\it Proof.}
By Lemma \ref{refinement-lemma}, Remark 4, there is a function $g$ such that
$\CausalState = g(\PrescientState)$ almost always.  But $H[f(X)] \leq H[X]$
(Eq.~(\ref{function-is-more-certain})) and so
\begin{equation}
H[\CausalState] = H[g(\PrescientState)] \leq H[\PrescientState] ~.
\end{equation}
but $\Cmu(\PrescientStateSet) = H[\PrescientState]$
(Def.~\ref{statistical-complexity-defined}).  QED.

{\it Remark 1.}
We have just established that no rival pattern, which is as good at predicting
the observations as the causal states, is any simpler, in the sense given by
Def.~\ref{statistical-complexity-defined}, than the causal states. (This
is the theorem of Ref. \cite{Inferring-stat-compl}.) Occam
therefore tells us that there is no reason not to use the causal states.  The
next theorem shows that causal states are uniquely optimal, and so that Occam's
Razor all but forces us to use them.

{\it Remark 2.}
Here it becomes important that we are trying to predict the whole of $\Future$
and not just some piece, $\FutureL$.  Suppose two histories $\past$ and
${\past}^{\prime}$ have the same conditional distribution for futures of
lengths up to $L$, but differing ones after that.  They would then belong to
different causal states.  An $\eta$-state that merged those two causal states,
however, would have just as much ability to predict $\FutureL$ as the causal
states.  More, these $\AlternateState$-states would be simpler, in the sense
that the uncertainty in the current state would be lower.  We conclude that
causal states are optimal, but for the hardest job---that of predicting
futures of all lengths.

{\it Remark 3.}
We have already seen (Thm.~\ref{optimal-prediction-theorem}, Remark 2) that
causal states are sufficient statistics for predicting futures of all lengths;
so are all prescient rivals.  A {\it minimal} sufficient statistic is one that
is a function of all other sufficient statistics \cite[p.~38]{Cover-and-Thomas}.  Since,
in the course of the proof of Thm.~\ref{minimality-theorem}, we have shown that
there is a function $g$ from any $\PrescientState$ to $\CausalState$, we have
also shown that causal states are minimal sufficient statistics.

We may now, as promised, define the {\em statistical complexity of a process}
\cite{Calculi-of-emergence,Inferring-stat-compl}.
\begin{definition}
{\bf (Statistical Complexity of a Process)}
The statistical complexity ``$\Cmu({\cal O})$'' of a process $\cal O$
is that of its causal states: $\Cmu ({\cal O}) \equiv \Cmu(\CausalStateSet)$.
\label{statistical-complexity-of-a-process}
\end{definition}
\addcontentsline{toc}{subsection}{\numberline{}Statistical Complexity of a Process}

Due to the minimality of causal states we see that the statistical complexity
measures the average amount of historical memory stored in the process.
Without the minimality theorem, this interpretation would not be possible,
since we could trivially elaborate internal states, while still generating the
same observed process.  $\Cmu$ for those states would grow without bound and so
be arbitrary and not a characteristic property of the process
\cite{JPC-semantics}.

\begin{theorem}
{\bf (Causal States Are Unique)}
For all prescient rivals $\PrescientStateSet$, if $\Cmu(\PrescientStateSet) =
\Cmu(\CausalStateSet)$, then there exists an invertible function between
$\PrescientStateSet$ and $\CausalStateSet$ that almost always preserves
equivalence of state: $\PrescientStateSet$ and $\eta$ are the same as
$\CausalStateSet$ and $\epsilon$, respectively, except on a set of histories of
measure $0$.
\label{uniqueness-theorem}
\end{theorem}
\addcontentsline{toc}{subsection}{\numberline{}Causal States Are Unique}

{\it Proof.}
From Lemma \ref{refinement-lemma}, we know that $\CausalState =
g(\PrescientState)$ almost always.  We now show that there is a function $f$
such that $\PrescientState = f(\CausalState)$ almost always, implying that $g =
{f}^{-1}$ and that $f$ is the desired relation between the two sets of states.
To do this, by Eq.~(\ref{functions-are-certain}) it is sufficient to show that
$H[\PrescientState|\CausalState] = 0$.  Now, it follows from an
information-theoretic identity (Eq.~(\ref{equality-of-mutual-information}))
that
\begin{equation}
H[\CausalState] - H[\CausalState|\PrescientState] = H[\PrescientState] - H[\PrescientState|\CausalState] ~.
\label{StateMutualInfo}
\end{equation}
Since, by Lemma \ref{refinement-lemma} $H[\CausalState|\PrescientState] = 0$,
both sides of Eq.~(\ref{StateMutualInfo}) are equal to $H[\CausalState]$.  But,
by hypothesis, $H[\PrescientState] = H[\CausalState]$.  Thus,
$H[\PrescientState|\CausalState] = 0$ and so there exists an $f$ such that
$\PrescientState = f(\CausalState)$ almost always.  We have then that
$f(g(\PrescientState)) = \PrescientState$ and $g(f(\CausalState)) =
\CausalState$, so $g = {f}^{-1}$.  This implies that $f$ preserves equivalence
of states almost always: for almost all $\past, \past^\prime \in \AllPasts$,
$\eta(\past) = \eta(\past^\prime)$ if and only if $\epsilon(\past) =
\epsilon({\past}^{\prime})$.  QED.

{\it Remark.}
As in the case of the Refinement Lemma \ref{refinement-lemma}, on which the
theorem is based, the measure-$0$ caveats seem unavoidable.  A rival that is as
predictive and as simple (in the sense of
Def.~\ref{statistical-complexity-defined}) as the causal states, can assign a
measure-$0$ set of histories to different states than the $\epsilon$-machine
does, but no more.  This makes sense: such a measure-$0$ set makes no
difference, since its members are never observed, by definition.  By the same
token, however, nothing prevents a minimal, prescient rival from disagreeing
with the $\epsilon$-machine on those histories.

\begin{theorem}
{\bf ($\epsilon$-Machines Are Minimally Stochastic)}
\cite{TDCS}
For all prescient rivals $\PrescientStateSet$,
\begin{equation}
H[\PrescientState^{\prime}|\PrescientState] \geq H[\CausalState^{\prime}|\CausalState] ~,
\end{equation}
where $\CausalState^{\prime}$ and $\PrescientState^{\prime}$ are the next
causal state of the process and the next $\eta$-state, respectively.
\label{minimal-stochasticity-theorem}
\end{theorem}
\addcontentsline{toc}{subsection}{\numberline{}$\epsilon$-Machines Are
Minimally Stochastic}

{\it Proof.}
From Lemma \ref{automatism-lemma}, $\CausalState^\prime$ is fixed by
$\CausalState$ and $\Future^1$ together, thus
$H[\CausalState^\prime|\CausalState, \Future^1] = 0$ by
Eq.~(\ref{functions-are-certain}).  Therefore, from the chain rule for entropies
Eq.~(\ref{conditional-chain-rule}),
\begin{eqnarray}
H[{\Future}^{1}|\CausalState] & = & H[\CausalState^{\prime}, {\Future}^{1}|\CausalState] ~.
\label{conditional-entropy-of-next-symbol-equals-entropy-of-next-causal-state}
\end{eqnarray}
We have no result like the Determinism Lemma \ref{automatism-lemma} for the
rival states $\PrescientState$, but entropies are always non-negative:
$H[\PrescientState^{\prime}|\PrescientState, {\Future}^{1}] \geq 0$.  Since for
all $L$, $H[\FutureL|\PrescientState] = H[\FutureL|\CausalState]$ by the
definition, Def. (\ref{PrescientRivals}), of prescient rivals,
$H[{\Future}^{1}|\PrescientState] = H[{\Future}^{1}|\CausalState]$.  Now we
apply the chain rule again,
\begin{eqnarray}
H[\PrescientState^{\prime},{\Future}^{1}|\PrescientState] & = &
H[{\Future}^{1}|{\PrescientState}] + H[\PrescientState^{\prime}|{\Future}^{1},
\PrescientState] \\
& \geq & H[{\Future}^{1}|\PrescientState] \\
\label{conditional-entropy-of-next-symbol}
& = & H[{\Future}^{1}|\CausalState] \\
\label{conditional-entropy-of-next-symbol-and-state}
& = & H[\CausalState^{\prime}, {\Future}^{1}|\CausalState] \\
& = & H[\CausalState^{\prime}|\CausalState] + H[{\Future}^{1}|\CausalState^{\prime}, \CausalState] ~.
\label{FirstExpansion}
\end{eqnarray}
In going from Eq.~(\ref{conditional-entropy-of-next-symbol}) to
Eq.~(\ref{conditional-entropy-of-next-symbol-and-state}) we have used
Eq.~(\ref{conditional-entropy-of-next-symbol-equals-entropy-of-next-causal-state}),
and in the last step we have used the chain rule once more.

Using the chain rule one last time, we have
\begin{equation}
H[\PrescientState^{\prime},{\Future}^{1}|\PrescientState]
  = H[{\PrescientState}^{\prime}|\PrescientState] + H[{\Future}^{1}|\PrescientState^{\prime}, \PrescientState] ~.
\label{SecondExpansion}
\end{equation}
Putting these expansions, Eqs.~(\ref{FirstExpansion}) and
(\ref{SecondExpansion}), together we get
\begin{eqnarray}
H[\PrescientState^{\prime}|\PrescientState] + H[{\Future}^{1}|{\PrescientState}^{\prime}, \PrescientState]
    & \geq & H[\CausalState^{\prime}|\CausalState] + H[{\Future}^{1}|\CausalState^{\prime}, \CausalState] \\
\nonumber
  H[\PrescientState^{\prime}|\PrescientState] - H[{\CausalState}^{\prime}|\CausalState]
	& \geq & H[{\Future}^{1}|\CausalState^{\prime}, \CausalState]
	  - H[{\Future}^{1}|\PrescientState^{\prime}, \PrescientState] ~.
\end{eqnarray}
From Lemma \ref{refinement-lemma}, we know that $\CausalState =
g(\PrescientState)$, so there is another function ${g}^{\prime}$ from ordered
pairs of $\eta$-states to ordered pairs of causal states:
$(\CausalState^{\prime}, \CausalState) = {g}^{\prime}(\PrescientState^{\prime},
\PrescientState)$.  Therefore, Eq.~(\ref{conditioning-on-function}) implies
\begin{equation}
H[\Future^1 | \CausalState^\prime, \CausalState] \geq H[\Future^1 | \PrescientState^\prime, \PrescientState] ~.
\end{equation}
And so, we have that
\begin{eqnarray}
\nonumber
H[{\Future}^{1}|\CausalState^{\prime}, \CausalState] - H[{\Future}^{1}|\PrescientState^{\prime}, \PrescientState]
  & \geq & 0 \\ \nonumber
H[\PrescientState^{\prime}|\PrescientState] - H[\CausalState^{\prime}|\CausalState] & \geq & 0 \\
H[\PrescientState^{\prime}|\PrescientState] & \geq & H[\CausalState^{\prime}|\CausalState] ~.
\end{eqnarray}
QED.

{\it Remark.}
What this theorem says is that there is no more uncertainty in transitions
between causal states, than there is in the transitions between any other kind
of prescient effective states.  In other words, the causal states approach as
closely to perfect determinism---in the usual physical,
non-computation-theoretic sense---as any rival that is as good at predicting
the future.  This sort of internal determinism has long been held to be a
desideratum of scientific models \cite{Bernard}.

\section{Bounds}
\label{Bounds}

In this section we develop bounds between measures of structural complexity and
entropy derived from $\epsilon$-machines and those from ergodic and information
theories, which are perhaps more familiar.

\begin{definition}[Excess Entropy]
The {\it excess entropy} $\EE$ of a process is the mutual information between
its semi-infinite past and its semi-infinite future:
\begin{equation}
\EE \equiv I[\Future;\Past] ~.
\end{equation}
\label{ExcessEntropyDefn}
\end{definition}
\addcontentsline{toc}{subsection}{\numberline{}Excess Entropy}

The excess entropy is a frequently-used measure of the complexity of stochastic
processes and appears under a variety of names; e.g., ``predictive
information'', ``stored information'', ``effective measure complexity'', and
so on \cite{JPC-Packard-noisy-chaos,Shaw-dripping,Grassberger-1986,Lindgren-Nordahl-1988,Li-complexity-vs-entropy,Arnold-info-theory-phase-trans,Bialek-predictive-info}.  $\EE$
measures the amount of {\em apparent} information stored in the observed
behavior about the past.  As we now establish, $\EE$ is not, in general, the
amount of memory that the process stores {\em internally} about its past; a
quantity measured by $\Cmu$.

\begin{theorem}[The Bounds of Excess]
The statistical complexity $\Cmu$ bounds the excess entropy $\EE$:
\begin{equation}
{\EE} \leq \Cmu ~,
\end{equation}
with equality if and only if $H[\CausalState|\Future] = 0$.
\label{bounded-excess-theorem}
\end{theorem}
\addcontentsline{toc}{subsection}{\numberline{}The Bounds of Excess}

{\it Proof.}
${\EE} = I[\Future; \Past] = H[\Future] - H[\Future| \Past]$ and, by
the construction of causal states, $H[\Future|\Past] = H[\Future|\CausalState]$, so
\begin{equation}
{\EE} = H[\Future] - H[\Future|\CausalState] = I[\Future; \CausalState] ~.
\end{equation}
Thus, since the mutual information between two variables is never larger than
the self-information of either one of them
(Eq.~(\ref{no-more-mutual-information-than-self-information})), ${\EE} \leq
H[\CausalState] = \Cmu$, with equality if and only if $H[\CausalState|\Future]
= 0$.  QED.

{\it Remark 1.}
Note that we have invoked $H[\Future]$, not $H[\FutureL]$, but only while
subtracting off quantities like $H[\Future|\Past]$.  We need not worry,
therefore, about the existence of a finite $\LLimit$ limit for
$H[\FutureL]$, just that of a finite $\LLimit$ limit for
$I[\FutureL;\Past]$ and $I[\FutureL;\CausalState]$.  There are many elementary
cases (e.g., the fair coin process) where the latter limits exist while the
former do not.

{\it Remark 2.}
At first glance, it is tempting to see $\EE$ as the amount of information
stored in a process.  As Thm.~\ref{bounded-excess-theorem} shows, this
temptation should be resisted.  $\EE$ is only a lower bound on the true amount
of information the process stores about its history, namely $\Cmu$.  We can,
however, say that $\EE$ measures the {\em apparent} information in the process,
since it is defined directly in terms of observed sequences and not in terms of
hidden, intrinsic states, as $\Cmu$ is.

{\it Remark 3.}
Perhaps another way to describe what $\EE$ measures is to note that, by its
implicit assumption of block-Markovian structure, it takes sequence-blocks as
states.  But even for the class of block-Markovian sources, for which such an
assumption is appropriate, excess entropy and statistical complexity measure
different kinds of information storage.  Refs.~\cite{DNCO} and \cite{JPC-DPF-stat-compl-of-1d-spin}
showed that in the case of one-dimensional range-$R$ spin systems, or any other
block-Markovian source where block configurations are isomorphic to causal
states:
\begin{equation}
\Cmu = \EE + R \hmu ~,
\end{equation}
for finite $R$.  Only for zero-entropy-rate block-Markovian sources will the
excess entropy, a quantity estimated directly from sequence blocks, equal the
statistical complexity, the amount of memory stored in the process.  Examples
of such sources include periodic processes, for which we have
$\Cmu = \EE = \log_2 p$, where $p$ is the period.

\begin{corollary}
For all prescient rivals $\PrescientStateSet$,
\begin{equation}
\EE \leq H[\PrescientState] ~.
\end{equation}
\label{PrescientRivalBoundsExcessEntropy}
\end{corollary}

{\it Proof.}
This follows directly from Thm.~\ref{minimality-theorem}, since
$H[\PrescientState] \geq \Cmu$.  QED.

\begin{lemma}
{\bf (Conditioning Does Not Affect Entropy Rate)}
For all prescient rivals $\PrescientStateSet$,
\begin{equation}
h[\Future] = h[\Future|\PrescientState] ~,
\end{equation}
where the entropy rate $h[\Future]$ and the conditional entropy rate $h[\Future|\PrescientState]$ were defined in Eq.~(\ref{EntropyRateBlockDefn}) and Eq.~(\ref{ConditionalEntropyRateDefn}), respectively.
\label{all-dead-in-the-long-run-lemma}
\end{lemma}
\addcontentsline{toc}{subsection}{\numberline{}Conditioning Does Not Affect
Entropy Rate}

{\it Proof.}
From Thm.~\ref{bounded-excess-theorem} and its Corollary
\ref{PrescientRivalBoundsExcessEntropy}, we have
\begin{equation}
\lim_{\LLimit}{ \left( H[\FutureL] - H[\FutureL|\PrescientState] \right) }
  \leq \lim_{\LLimit}{H[\PrescientState]} ~,
\end{equation}
or,
\begin{equation}
\lim_{\LLimit}{H[\FutureL] - H[\FutureL|\PrescientState] \over L} \leq
\lim_{\LLimit}{H[\PrescientState] \over L} ~.
\end{equation}
Since, by Eq.~(\ref{conditioning-reduces-entropy}), $H[\FutureL] -
H[\FutureL|\PrescientState] \geq 0$, we have
\begin{equation}
h[\Future] - h[\Future|\PrescientState] = 0 ~.
\end{equation}
QED.

{\it Remark.}
Forcing the process into a certain state $\PrescientState = \prescientstate$ is
akin to applying a controller, once.  But in the infinite-entropy case,
$H[\FutureL] \rightarrow_{\LLimit} \infty$, with which we are
concerned, the future could contain (or consist of) an infinite sequence of
disturbances.  In the face of this ``grand disturbance'', the effects
of the finite control are simply washed out.

Another way of viewing this is to reflect on the fact that $h[\Future]$
accounts for the effects of all the dependencies between all the parts of the
entire semi-infinite future.  This, owing to the time-translation invariance of
stationarity, is equivalent to taking account of all the dependencies in the
entire process, including those between past and future.  But these are what is
captured by $h[\Future|\PrescientState]$.  It is not that conditioning on
$\AlternateState$ fails to reduce our uncertainty about the future; it does so,
for all finite times, and conditioning on $\CausalState$ achieves the maximum
possible reduction in uncertainty.  Rather, the lemma asserts that such
conditioning cannot effect the asymptotic rate at which such uncertainty grows
with time.

\begin{theorem}[Control Theorem]
Given a class $\PrescientStateSet$ of prescient rivals,
\begin{equation}
H[S] - h[\Future|\PrescientState] \leq \Cmu ~,
\end{equation}
where $H[S]$ is the entropy of a single symbol from the observable stochastic
process.
\label{control-theorem}
\end{theorem}
\addcontentsline{toc}{subsection}{\numberline{}Control Theorem}

{\it Proof.}
As is well known (Ref.~\cite[Thm.~4.2.1, p.~64]{Cover-and-Thomas}), for any stationary
stochastic process,
\begin{equation}
\lim_{\LLimit}{H[\FutureL] \over L} =
\lim_{\LLimit}{H[S_L|\FutureLLessOne]} ~.
\label{EquivalenceOfEntropyRateDefns}
\end{equation}
Moreover, the limits always exist.  Up to this point, we have defined
$h[\Future]$ in the manner of the left-hand side; recall
Eq.~(\ref{EntropyRateBlockDefn}).  It will be convenient in the following to
use that of the right-hand side.

From the definition of conditional entropy, we have
\begin{eqnarray}
\nonumber
H[\PastL] & = & H[\LastObservable|\PastLLessOne] + H[\PastLLessOne]\\
& = & H[\PastLLessOne|\LastObservable] + H[\LastObservable] ~.
\label{two-sums-for-the-past}
\end{eqnarray}
So we can express the entropy of the last observable the process generated
before the present as
\begin{eqnarray}
\label{first-expression-for-the-last-symbol}
H[\LastObservable] & = & H[\PastL] - H[\PastLLessOne|\LastObservable] \\
\label{second-expression-for-the-last-symbol}
  & = & H[\LastObservable|\PastLLessOne] + H[\PastLLessOne]
    - H[\PastLLessOne|\LastObservable] \\
  & = & H[\LastObservable|\PastLLessOne]+I[\PastLLessOne;\LastObservable] ~.
\end{eqnarray}
We go from Eq.~(\ref{first-expression-for-the-last-symbol}) to
Eq.~(\ref{second-expression-for-the-last-symbol}) by substituting the first RHS
of Eq.~(\ref{two-sums-for-the-past}) for $H[\PastL]$.

Taking the $\LLimit$ limit has no effect on the LHS,
\begin{equation}
H[\LastObservable] =
  \lim_{\LLimit} \left( H[\LastObservable|\PastLLessOne]
  + I[\PastLLessOne; \LastObservable] \right) ~.
\label{SingleSymbolEntropyPieces}
\end{equation}
Since the process is stationary, we can move the first term in the limit
forward to $H[{S}_{L}|\FutureLLessOne]$.  This limit is $h[\Future]$, by
Eq.~(\ref{EquivalenceOfEntropyRateDefns}).  Furthermore, because of
stationarity, $H[\LastObservable] = H[\NextObservable] = H[S]$.  Shifting the
entropy rate $h[\Future]$ to the LHS of Eq.~(\ref{SingleSymbolEntropyPieces})
and appealing to time-translation once again, we have
\begin{eqnarray}
H[S] - h[\Future] & = & \lim_{\LLimit}{I[\PastLLessOne;\LastObservable]}\\
	& = & I[\Past;\NextObservable]\\
	& = & H[\NextObservable] - H[\NextObservable|\Past]\\
	& = & H[\NextObservable] - H[\NextObservable|\CausalState]\\
	& = & I[\NextObservable;\CausalState]\\
	& \leq & H[\CausalState] = \Cmu ~,
\end{eqnarray}
where the last inequality comes from
Eq.~(\ref{no-more-mutual-information-than-self-information}).  QED.

{\it Remark 1.}
The Control Theorem is inspired by, and is a version of, Ashby's
{\em law of requisite variety} \cite[ch.~11]{Ashby-I-to-C}. This states that
applying a controller can reduce the uncertainty in the controlled variable
by at most the entropy of the control variable. (This result has recently
been rediscovered in Ref.~\cite{Touchette-Lloyd}.)  Thinking of the controlling
variable as the causal state, we have here a limitation on the controller's
ability to reduce the entropy {\it rate}.

{\it Remark 2.}
This is the only result so far where the difference between the finite-$L$
and the infinite-$L$ cases is important.  For the analogous result in the
finite case, see App.~\ref{FiniteEntropyInfiniteFuture}, Thm.~\ref{finite-control-theorem}.

{\it Remark 3.}
By applying Thm.~\ref{minimality-theorem} and Lemma
\ref{all-dead-in-the-long-run-lemma}, we could go from the theorem as it stands
to $H[S] - h[\Future|\PrescientState] \leq H[\PrescientState]$.  This has a
pleasing appearance of symmetry to it, but is actually a weaker limit on the
strength of the pattern or, equivalently, on the amount of control that fixing
the causal state (or one of its rivals) can exert.

\section{Concluding Remarks}
\label{Conclusion}

\subsection{Discussion}

Let's review, informally, what we have shown.  We began with questions about
the nature of patterns and about pattern discovery.  Our examination of these
issues lead us to want a way of describing patterns that was at once algebraic,
computational, intrinsically probabilistic, and causal.  We then defined
patterns in ensembles, in a very general and abstract sense, as equivalence
classes of histories, or sets of hidden states, used for prediction.  We
defined the strength of such patterns (by their forecasting ability or
prescience) and their statistical complexity (by the entropy of the states or
the amount of information retained by the process about its history).  We
showed that there was a limit on how strong such patterns could get for each
particular process, given by the predictive ability of the entire past. In this
way, we narrowed our goal to finding a predictor of maximum strength and
minimum complexity.

Optimal prediction led us to the equivalence relation
$\CausalEquivalence$ and the function $\epsilon$ and so to representing
patterns by causal states and their transitions---the $\epsilon$-machine.  Our
first theorem showed that the causal states are maximally prescient; our
second, that they are the simplest way of representing the pattern of maximum
strength; our third theorem, that they are unique in having this double
optimality.  Further results showed that $\epsilon$-machines are the least
stochastic way of capturing maximum-strength patterns and emphasized the need
to employ the efficacious but hidden states of the process, rather than just
its gross observables, such as sequence blocks.

Why are $\epsilon$-machine states causal? First, $\epsilon$-machine
architecture (say, as given by its semi-group algebra) delineates the
dependency between the morphs $\Prob(\Future|\Past)$, considered as events in
which each new symbol determines the succeeding morph.  Thus, if state $B$
follows state $A$ then $A$ is a cause of $B$ and $B$ is an effect of $A$.
Second, $\epsilon$-machine minimality guarantees that there are no other events
that intervene to render $A$ and $B$ independent \cite{JPC-semantics}.

The $\epsilon$-machine is thus a causal representation of all the patterns in
the process.  It is maximally predictive and minimally complex.  It is at once
computational, since it shows how the process stores information (in the causal
states) and transforms that information (in the state-to-state transitions),
and algebraic (for details on which see App.~\ref{SemiGroups}).  It can be
analytically calculated from given distributions and systematically approached
from empirical data.  It satisfies the basic constraints laid out in
Sec.~\ref{Desiderata}.

These comments suggest that computational mechanics and $\epsilon$-machines are
related or may be of interest to a number of fields.  Time series analysis,
decision theory, machine learning, and universal coding theory explicitly or
implicitly require models of observed processes.  The theories of stochastic
processes, formal languages and computation, and of measures of physical
complexity are all concerned with representations of processes---concerns which
also arise in the design of novel forms of computing devices.  Very often the
motivations of these fields are far removed from computational mechanics.  But
it is useful, if only by way of contrast, to touch briefly on these areas and
highlight one or several connections with computational mechanics, and we do so
in App.~\ref{compare-and-contrast}.

\subsection{Limitations of the Current Results}
\label{ThingsThatAreNotYetTheorems}

Let's catalogue the restrictive assumptions we made at the beginning and that
were used by our development.
\begin{enumerate}
\item We know exact joint probabilities over sequence blocks of all lengths for
a process.
\item The observed process takes on discrete values.
\item The process is discrete in time.
\item The process is a pure time series; e.g., without spatial extent.
\item The observed process is stationary.
\item Prediction can only be based on the process's past, not on any outside
source of information.
\end{enumerate}
The question arises, Can any be relaxed without much trouble?

One way to lift the first limitation is to develop a statistical error theory
for $\epsilon$-machine inference that indicates, say, how much data is required
to attain a given level of confidence in an $\epsilon$-machine with a given
number of causal states.  This program is underway and, given its initial
progress, we describe several issues in more detail in the next section.

The second limitation probably can be addressed, but with a corresponding
increase in mathematical sophistication.  The information-theoretic quantities
we have used are also defined for continuous random variables.  It is likely
that many of the results carry over to the continuous setting.

The third limitation also looks similarly solvable, since continuous-time
stochastic process theory is moderately well developed.  This may involve
sophisticated probability theory or functional analysis.

As for the fourth limitation, there already exist tricks to make spatially
extended systems look like time series.  Essentially, one looks at all the
paths through space-time, treating each one as if it were a time series.  While
this works well for data compression\cite{Lempel-Ziv-two-d}, it is not yet clear whether
it will be entirely satisfactory for capturing structure \cite{DPF-thesis}.
More work needs to be done on this subject.

It is unclear at this time how to relax the assumption of stationarity.  One
can formally extend most of the results in this paper to non-stationary
processes without much trouble.  It is, however, unclear how much substantive
content these extensions have and, in any case, a systematic classification of
non-stationary processes is (at best) in its infant stages.

Finally, one might say that the last restriction is a positive {\it feature}
when it comes to thinking about patterns and the intrinsic structure of a
process.  ``Pattern'' is a vague word, of course, but even in ordinary usage it
is only supposed to involve things {\it inside} the process, not the rest of
the universe.  Given two copies of a document, the contents of one copy can
be predicted with an enviable degree of accuracy by looking at the other
copy.  This tells us that they share a common structure, but says absolutely
nothing about what that pattern is, since it is just as true of well-written
and tightly-argued scientific papers (which presumably are highly organized)
as it is of monkey-at-keyboard pieces of gibberish (which definitely are
not).

\subsection{Conclusions and Directions for Future Work}

Computational mechanics aims to understand the nature of patterns and pattern
discovery.  We hope that the foregoing development has convinced the reader
that we are neither being rash when we say that we have laid a foundation for
those projects, nor that we are being flippant when we say that patterns are
what $\epsilon$-machines represent and that we discover them by
$\epsilon$-machine reconstruction.  We would like to close by marking out two
broad avenues for future work.

First, consider the mathematics of $\epsilon$-machines themselves.  We have
just mentioned possible extensions in the form of lifting assumptions made in
this development, but there are many other ways to go.  A number of
measure-theoretic issues relating to the definition of causal states (omitted
here for brevity) deserve careful treatment, along the lines of Ref.
\cite{Upper-thesis}. It would be helpful to have a good understanding of the
measurement-resolution scaling properties of $\epsilon$-machines for
continuous-state processes, and of their relation to such ideas in automata
theory as the Krohn-Rhodes decomposition \cite{Rhodes-book}. Anyone who manages to
absorb Volume II of Ref.~\cite{Principia} would probably be in a position to
answer interesting questions about the structures that processes preserve,
perhaps even to give a purely relation-theoretic account of $\epsilon$-machines.
We have alluded in a number of places to the trade-off between prescience and
complexity.  For a given process there is presumably a sequence of optimal
machines connecting the one-state, zero-complexity machine with minimal
prescience to the $\epsilon$-machine.  Each member of the path is the minimal
machine for a certain degree of prescience; it would be very interesting to
know what, if anything, we can say in general about the shape of this
``prediction frontier''.

Second, there is $\epsilon$-machine reconstruction, an activity about which we
have said next to nothing.  As we mentioned above
(p.~\pageref{ReconstructionDefn}), there are already several algorithms for
reconstructing machines from data, even ``on-line'' ones.  It is fairly evident
that these algorithms will find the true machine in the limit of infinite time
and infinite data.  What is needed is an understanding of the {\it error
statistics} \cite{Mayo-error} of different reconstruction procedures of the kinds
of mistakes these procedures make and the probabilities with which they make
them.  Ideally, we want to find ``confidence regions'' for the products of
reconstruction.  The aim is to calculate (i) the probabilities of different
degrees of reconstruction error for a given volume of data, (ii) the amount of
data needed to be confident of a fixed bound on the error, or (iii) the rates
at which different reconstruction procedures converge on the
$\epsilon$-machine.  So far, an analytical theory has been developed that
predicts the average number of estimated causal states as a function of the
amount of data used when reconstructing certain kinds of processes
\cite{JPC-and-Chris-Douglas}.  Once we possess a more complete theory of statistical
inference for $\epsilon$-machines, analogous perhaps to what already exists in
computational learning theory, we will be in a position to begin analyzing,
sensibly and rigorously, the multitude of intriguing patterns and
information-processing structures the natural world presents.

\section*{Acknowledgments}

We thank Dave Albers, Dave Feldman, Jon Fetter, Rob Haslinger, Wim Hordijk,
Amihan Huesmann, Cris Moore, Mitch Porter, Erik van Nimwegen, and Karl Young
for advice on the manuscript; and the participants in the 1998 SFI Complex
Systems Summer School, the Madison probability seminar, the Madison Physics
Department's graduate student mini-colloquium, and the Ann Arbor Complex
Systems seminar for numerous helpful comments on earlier versions of these
results.  This work was supported at the Santa Fe Institute under the
Computation, Dynamics, and Inference Program via ONR grant N00014-95-1-0975,
NSF grant PHY-9970158, and DARPA contract F30602-00-2-0583.

\begin{appendix}

\section[Information-Theoretic Formul{\ae}]{Information-Theoretic Formul{\AE}}
\label{InfoTheoryFormulae}

The following formul{\ae} prove useful in the development.  They are
relatively intuitive, given our interpretation, and they can all be proved with
little more than straight algebra; see Ref.~\cite[ch.~2]{Cover-and-Thomas}.  Below, $f$
is a function.
\begin{eqnarray}
H[X, Y]       & = & H[X] + H[Y|X] \label{ChainRule} \\
H[X, Y]       & \geq & H[X] \\
H[X, Y]       & \leq & H[X] + H[Y] \label{IndepBound} \\
H[X|Y]        & \leq & H[X] \label{conditioning-reduces-entropy} \\
H[X|Y] = H[X] & {\rm iff} & X {\rm is\ independent\ of\ } Y \\
H[X, Y|Z]     & = & H[X|Z] + H[Y|X, Z] \label{conditional-chain-rule} \\
H[X, Y|Z]     & \geq & H[X|Z] \\
H[X] - H[X|Y] & = & H[Y] - H[Y|X] \label{equality-of-mutual-information} \\
I[X;Y] & \leq & H[X] \label{no-more-mutual-information-than-self-information} \\
I[X;Y] = H[X] & {\rm iff} & H[X|Y] = 0 \\
H[f(X)]       & \leq & H[X] \label{function-is-more-certain} \\
H[X|Y] = 0    & {\rm iff} &  X = f(Y) \label{functions-are-certain} \\
H[f(X)|Y]     & \leq & H[X|Y] \label{conditioned-function} \\
H[X|f(Y)]     & \geq & H[X|Y] \label{conditioning-on-function}
\end{eqnarray}
Eqs.~(\ref{ChainRule}) and (\ref{conditional-chain-rule}) are called the {\em
chain rules} for entropies.  Strictly speaking, the right hand side of
Eq.~(\ref{functions-are-certain}) should read ``for each $y$,
$\Prob(X=x|Y=y) > 0$ for one and only one $x$''.

\section{The Equivalence Relation that Induces Causal States}
\label{ReviewEquivalenceRelation}

Any relation that is reflexive, symmetric, and transitive is an {\it
equivalence relation}.

Consider the set $\AllPasts$ of all past sequences, of any length:
\begin{equation}
\AllPasts = \{ \past^L = s_{L-1} \cdots s_{-1} : \;
s_i \in {\cal A}, \; L \in {\Bbb Z}^{+} \} \; .
\label{PastSetDefn}
\end{equation}
Recall that $\past^0 = \lambda$, the empty string.  We define the
relation $\CausalEquivalence$ over $\AllPasts$ by
\begin{equation}
{\stackrel{\leftarrow}{s_{i}}}^K ~\CausalEquivalence~
  {\stackrel{\leftarrow}{s_{j}}}^L \; \Leftrightarrow \;
  \Prob(\Future | {\stackrel{\leftarrow}{s_{i}}}^K) =
  \Prob(\Future | {\stackrel{\leftarrow}{s_{j}}}^L) \;,
\end{equation}
for all semi-infinite $\Future = s_0 s_1 s_2 \cdots$, where $K, L \in {\Bbb Z}^{+}$.  Here we show that $\sim_\epsilon$ is an equivalence relation by
reviewing the basic properties of relations, equivalence classes, and
partitions.  (The proof details are straightforward and are not included.
See Ref.~\cite{Lidl}.)  We will drop the length variables $K$ and $L$ and
denote by $\past, \PastPrime, \PastDblPrime \in \,\AllPasts$ members of any
length in the set $\AllPasts$ of Eq.~(\ref{PastSetDefn}).

First, $\CausalEquivalence$ is a {\em relation} on $\AllPasts$ since we can represent it as a
subset of the Cartesian product
\begin{equation}
\AllPasts \times \AllPasts = \{ (\past,\PastPrime)
: \past,\PastPrime \in \AllPasts \}\;.
\end{equation}

Second, the relation $\CausalEquivalence$ is an {\em equivalence relation} on $\AllPasts$
since it is
\begin{enumerate}
\item reflexive: $\past \CausalEquivalence \past$,
for all $\past \in \AllPasts$;
\item symmetric: $\past \CausalEquivalence \PastPrime \Rightarrow
\PastPrime \CausalEquivalence \past$; and
\item transitive: $\past \CausalEquivalence \PastPrime
\;{\rm and}\; \PastPrime \CausalEquivalence \PastDblPrime
\Rightarrow \past \CausalEquivalence \PastDblPrime$.
\end{enumerate}

Third, if $\past \in \AllPasts$, the {\em equivalence class} of $\past$ is
\begin{equation}
[ \past ] = \{ \PastPrime \in \AllPasts :
\PastPrime \CausalEquivalence \past \} \; .
\end{equation}
The set of all equivalence classes in $\AllPasts$ is denoted $\AllPasts /
\CausalEquivalence$ and is called the {\em factor set} of $\AllPasts$ with
respect to $\CausalEquivalence$.  In Sec.~\ref{DefnCausalStatesEMs} we called
the individual equivalence classes {\em causal states} $\CausalState_i$ and
denoted the set of causal states $\CausalStateSet = \{ \CausalState_i : i = 0,
1, \ldots, k-1 \}$.  That is, $\CausalStateSet = \AllPasts /
\CausalEquivalence$.  (We noted in the main development that the cardinality $k
= | \CausalStateSet |$ of causal states may or may not be finite.)

Finally, we list several basic properties of the causal-state equivalence
classes.
\begin{enumerate}
\item $\bigcup_{\past \in \AllPasts} [ \past ] = \AllPasts \; $.
\item $\bigcup_{i=0}^{k-1} \CausalState_i = \AllPasts \;$.
\item $[\past] = [ \PastPrime ] \Leftrightarrow \,\past \CausalEquivalence \PastPrime \;$.
\item If $\past, \PastPrime \in \AllPasts$, either
        \begin{enumerate}
        \item $[\past] \bigcap \,[ \PastPrime ] = \emptyset$ or
        \item $[\past] = [ \PastPrime ] \;$.
        \end{enumerate}
\item The causal states $\CausalStateSet$ are a partition of $\AllPasts$.
That is,
        \begin{enumerate}
        \item $\CausalState_i \neq \emptyset$ for each $i$,
        \item $\bigcup_{i=0}^{k-1} \CausalState_i = \AllPasts$, and
        \item $\CausalState_i \cap \CausalState_j = \emptyset$ for
                all $i \neq j$.
        \end{enumerate}
\end{enumerate}

We denote the start state with $\CausalState_0$.  The start state is the causal
state associated with $\past = \lambda$.  That is, $\CausalState_0 = [\lambda]$.

\section{Time Reversal}
\label{TimeReversal}

The definitions and properties of the causal states obtained by scanning
sequences in the opposite direction, i.e., the causal states $\FutureSet /
\CausalEquivalence$, follow similarly to those derived just above in App.
\ref{ReviewEquivalenceRelation}.  In general, $\AllPasts / \CausalEquivalence \;
\neq \; \FutureSet / \CausalEquivalence$.  That is, {\em past} causal states are
not necessarily the same as {\em future} causal states; past and future
morphs can differ; unlike entropy rate \cite{TDCS}, past
and future statistical complexities need not be equal:
$\stackrel{\leftarrow}{\Cmu} \neq \stackrel{\rightarrow}{\Cmu}$; and so on.
The presence or lack of this type of time-reversal symmetry, as reflected in
these inequalities, is a fundamental property of a process.

\section{$\epsilon$-Machines are Monoids}
\label{SemiGroups}

A {\em semi-group} is a set of elements closed under an associative binary
operator, but without a guarantee that every, or indeed any, element has an
inverse \cite{Ljapin-semigroups}.  A {\em monoid} is a semi-group with an identity
element.  Thus, semi-groups and monoids are generalizations of groups.  Just as
the algebraic structure of a group is generally interpreted as a symmetry, we
propose to interpret the algebraic structure of a semi-group as a {\em
generalized} symmetry. The distinction between monoids and other semi-groups
becomes important here: only semi-groups with an identity element---i.e.,
monoids---can contain subsets that are groups and so represent
conventional symmetries.

We claim that the transformations that concatenate strings of symbols from
${\cal A}$ onto other such strings form a semi-group $G$, the generators of
which are the transformations that concatenate the elements of ${\cal A}$.  The
identity element is to be provided by concatenating the null symbol $\lambda$.
The concatenation of string $t$ onto the string $s$ is forbidden if
and only if strings of the form $st$ have probability zero in a process.  All
such concatenations are to be realized by a single semi-group element denoted
$\emptyset$.  Since if $\Prob(st) = 0$, then $\Prob(stu)=\Prob(ust)=0$ for any
string $u$, we require that $\emptyset g = g \emptyset = \emptyset$ for all $g
\in G$.  Can we provide a representation of this semi-group?

Recall that, from our definition of the labeled transition probabilities,
${T}^{(\lambda)}_{ij} = \delta_{ij}$.  Thus, ${\bf T}^{(\lambda)}$ is an
identity element.  This suggests using the labeled transition matrices to form a
matrix representation of the semi-group.  Accordingly, first define
${U}^{(s)}_{ij}$ by setting ${U}^{(s)}_{ij} = 0$ when ${T}^{(s)}_{ij} = 0$ and
${U}^{(s)}_{ij} = 1$ otherwise, to remove probabilities. Then define the set
of matrices
$\bf{U} = \{ {\bf T}^{(\lambda)} \}
\bigcup \{ {\bf{U}}^{(s)} ~, s \in {\cal A} \}$.
Finally, define $G$ as the set of
all matrices generated from the set $\mathbf{U}$ by recursive multiplication.
That is, an element $g$ of $G$ is
\begin{equation}
g^{(ab\ldots cd)}
  = {\mathbf{U}}^{(d)}{\mathbf{U}}^{(c)}\ldots {\mathbf{U}}^{(b)}{\mathbf{U}}^{(a)} ~,
\label{SemiGroupElements}
\end{equation}
where $a, b, \ldots c, d \in {\cal A}$.  Clearly, $G$ constitutes a semi-group
under matrix multiplication.  Moreover, ${g}^{(a\ldots bc)} = \mathbf{0}$ (the
all-zero matrix) if and only if, having emitted the symbols $a\ldots b$ in
order, we must arrive in a state from which it is impossible to emit the symbol
$c$.  That is, the zero-matrix $\mathbf{0}$ is generated if and only if the
concatenation of $c$ onto $a\ldots b$ is forbidden.  The element $\emptyset$ is
thus the all-zero matrix $\mathbf{0}$, which clearly satisfies the necessary
constraints.  This completes the proof of Proposition \ref{eMSemiGroup}.

We call the matrix representation---Eq.~(\ref{SemiGroupElements}) taken over
all words in ${\cal A}^k$---of $G$ the {\em semi-group machine} of the
$\epsilon$-machine $\{ \CausalStateSet , {\bf T} \}$.  See Ref.~\cite{Young-thesis}.

\section{Alternate Proof of the Refinement Lemma}
\label{AltProof}

The proof of Lemma \ref{refinement-lemma} carries through verbally, but we do
not wish to leave loop-holes.  Unfortunately, this means introducing two new
bits of mathematics.

First of all, we need the largest classes that are strictly homogeneous
(Def.~\ref{StrictHomogeneityDefn}) with respect to $\FutureL$ for fixed $L$;
these are, so to speak, truncations of the causal states.  Accordingly, we will
talk about $\CausalState^L$ and $\causalstate^L$, which are analogous to
$\CausalState$ and $\causalstate$.  We will also need to define the function
$\phi_{\causalstate\alternatestate}^L \equiv \Prob(\CausalState^L =
\causalstate^L | \AlternateState = \alternatestate)$.

Putting these together, for every $L$ we have
\begin{eqnarray}
H[ \FutureL | \AlternateState = \alternatestate] & = & H[\sum_{{\causalstate}^{L}}{{\phi}_{\causalstate\alternatestate}^{L}\Prob(\FutureL|{\CausalState}^{L} = {\causalstate}^{L})}] \\
& \geq & \sum_{{\causalstate}^{L}}{{\phi}_{\causalstate\alternatestate}^{L}H[\FutureL|{\CausalState}^{L} = {\causalstate}^{L}]} ~.
\end{eqnarray}
Thus,
\begin{eqnarray}
H[\FutureL & | & \AlternateState] = \sum_{\alternatestate}{\Prob(\AlternateState = \alternatestate) H[\FutureL|\AlternateState = \alternatestate]} \\
& \geq & \sum_{\rho}{\Prob(\AlternateState = \alternatestate) \sum_{{\causalstate}^{L}}{{\phi}_{\causalstate\alternatestate}^{L}
H[\FutureL|{\CausalState}^{L} = {\causalstate}^{L}]}} \\
& = & \sum_{{\causalstate}^{L}, \alternatestate}{\Prob(\AlternateState = \alternatestate) {\phi}_{\causalstate\alternatestate}^{L} H[\FutureL|{\CausalState}^{L} =
{\causalstate}^{L}]} \\
& = & \sum_{{\causalstate}^{L}, \alternatestate}{\Prob({\CausalState}^{L} = {\causalstate}^{L} \ProbAnd \AlternateState = \alternatestate)
H[\FutureL|{\CausalState}^{L} = {\causalstate}^{L}]} \\
& = & \sum_{{\causalstate}^{L}}{\Prob({\CausalState}^{L} = {\causalstate}^{L}) H[\Future|{\CausalState}^{L} =
{\causalstate}^{L}]} \\
& = & H[\FutureL|{\CausalState}^{L}] ~.
\end{eqnarray}
That is to say,
\begin{eqnarray}
H[\FutureL| \AlternateState] & \geq & H[\FutureL | {\CausalState}^{L}] ~,
\end{eqnarray}
with equality if and only if every ${\phi}_{\causalstate\alternatestate}^{L}$
is either $0$ or $1$.  Thus, if $H[\FutureL|\PrescientState] =
H[\Future|{\CausalState}^{L}]$, every $\prescientstate$ is entirely contained
within some ${\causalstate}^{L}$; except for possible subsets of measure $0$.
But if this is true for every $L$---which, in the case of a prescient rival
$\PrescientState$, it is---then every $\prescientstate$ is at least weakly
homogeneous (Def.~\ref{WeakHomogeneityDefn}) with respect to all
$\FutureL$.  Thus, by Lemma \ref{homogeneity-lemma}, all its members, except for
that same subset of measure $0$, belong to the same causal state.  QED.

\section{Finite Entropy for the Semi-Infinite Future}
\label{FiniteEntropyInfiniteFuture}

While cases where $H[\Future]$ is finite---more exactly, where $\lim_{L
\rightarrow \infty}{H[\FutureL]}$ exists and is finite---may be uninteresting
for information-theorists, they are of great interest to physicists, since
they correspond, among other things, to periodic and limit-cycle behaviors.
There are, however, only two substantial differences between what is true of
the infinite-entropy processes considered in the main body of the development
and the finite-entropy case.

First, we can simply replace statements of the form ``for all $L$,
$H[{\Future}^{L}]$ \ldots '' with $H[\Future]$.  For example, the optimal
prediction theorem (Thm.~\ref{optimal-prediction-theorem}) for finite-entropy
processes becomes for all $\AlternateStateSet$,
$H[\Future|\AlternateState] \geq H[\Future|\CausalState]$.  The details of the proofs are,
however, entirely analogous.

Second, we can prove a substantially stronger version of the control
theorem (Thm.~\ref{control-theorem}).

\begin{theorem}[The Finite-Control Theorem]
For all prescient rivals $\PrescientStateSet$,
\begin{equation}
H[\Future] - H[\Future|\PrescientState] \leq \Cmu ~.
\end{equation}
\label{finite-control-theorem}
\end{theorem}
\addcontentsline{toc}{subsection}{\numberline{}The Finite-Control Theorem}

{\it Proof.}
By a direct application of
Eq.~(\ref{no-more-mutual-information-than-self-information}) and the
definition of mutual information Eq.~(\ref{def-of-mutual-info}),
we have that
\begin{equation}
H[\Future] - H[\Future|\CausalState] \leq H[\CausalState] ~.
\end{equation}
But, by the definition of prescient rivals (Def.~\ref{PrescientRivals}),
$H[\Future|\CausalState] = H[\Future|\PrescientState]$, and, by definition,
$\Cmu = H[\CausalState]$.  Substituting equals for equals gives us the
theorem.  QED.

\section{Relations to Other Fields}
\label{compare-and-contrast}

\subsection{Time Series Modeling}
\label{TimeSeriesModeling}

The goal of time series modeling is to predict the future of a measurement
series on the basis of its past.  Broadly speaking, this can be divided into
two parts: identify equivalent pasts and then produce a prediction for each
class of equivalent pasts.  That is, we first pick a function $\eta: \AllPasts
\mapsto \AlternateStateSet$ and then pick another function $p:
\AlternateStateSet \mapsto \AllFutures$.  Of course, we can choose for the
range of $p$ futures of some finite length (length 1 is popular) or even choose
distributions over these.  While practical applications often demand a single
definite prediction---``You will meet a tall dark stranger'', there are obvious
advantages to predicting a distribution---``You have a $.95$ chance of meeting
a tall dark stranger and a $.05$ chance of meeting a tall familiar albino.''
Clearly, the best choice for $p$ is the actual conditional distribution of
futures for each $\alternatestate \in \AlternateStateSet$.  Given this, the
question becomes what the best $\AlternateStateSet$ is; i.e., What is the best
$\eta$?  At least in the case of trying to understand the whole of the
underlying process, we have shown that the best $\eta$ is, unambiguously,
$\epsilon$.  Thus, our discussion has implicitly subsumed that of traditional
time series modeling.

Computational mechanics---in its focus on letting the process speak for itself
through (possibly impoverished) measurements---follows the spirit that
motivated one approach to experimentally testing dynamical systems theory.
Specifically, it follows in spirit the methods of reconstructing ``geometry
from a time series'' introduced by Refs.~\cite{Geometry-from-a-time-series} and \cite{Takens-embedding}.  A
closer parallel is found, however, in later work on estimating minimal
equations of motion from data series \cite{JPC-equations-of-motion}.

\subsection{Decision-Theoretic Problems}

The classic focus of decision theory is ``rules of inductive behavior''
\cite{Neymman-first-course,Blackwell-Girshick,Luce-and-Raiffa}.  The problem is to chose functions from observed
data to courses of action that possess desirable properties.  This task has
obvious affinities to considering the properties of $\epsilon$ and its rivals
$\eta$.  We can go further and say that what we have done {\it is} consider a
decision problem, in which the available actions consist of predictions about
the future of the process.  The calculation of the optimum rule of behavior in
general faces formidable technicalities, such as providing an estimate of the
utility of every different course of action under every different hypothesis
about the relevant aspects of the world.  On the one hand, it is not hard to
concoct time-series tasks where the optimal rule of behavior does not use
$\epsilon$ at all.  On the other hand, if we simply aim to predict the process
indefinitely far into the future, then because the causal states are minimal
sufficient statistics for the distribution of futures
(Thm.~\ref{minimality-theorem}, Remark 4), the optimal rule of behavior will
use $\epsilon$.

\subsection{Stochastic Processes}

Clearly, the computational mechanics approach to patterns and pattern discovery
involves stochastic processes in an intimate and inextricable way.
Probabilists have, of course, long been interested in using
information-theoretic tools to analyze stochastic processes, particularly their
ergodic behavior \cite{Billingsley-ergodic-theory-and-info,Gelfand-Yaglom-info,Caines,Gray-entropy}.  There has also been
considerable work in the hidden Markov model and optimal prediction literatures
on inferring models of processes from data or from given distributions
\cite{Upper-thesis,Blackwell-identifiability,Ito-et-al-identifiability,Jaeger-operator-models,Algoet-universal-schemes}.  To the best of our knowledge,
however, these two approaches have not been previously combined.

Perhaps the closest approach to the spirit of computational mechanics in the
stochastic process literature is, surprisingly, the now-classical theory of
optimal prediction and filtering for stationary processes, developed by Wiener
and Kolmogorov \cite{Kolmogorov-interpolation-extrapolation,Wiener-time-series,Wiener-nonlinear,Wiener-cybernetics}.  The two theories share the
use of information-theoretic notions, the unification of prediction and
structure, and the conviction that ``the statistical mechanics of time series''
is a ``field in which conditions are very remote from those of the statistical
mechanics of heat engines and which is thus very well suited to serve as a
model of what happens in the living organism'' \cite[p.~59]{Wiener-cybernetics}.  So far as
we have been able to learn, however, no one has ever used this theory to
explicitly identify causal states and causal structure, leaving these implicit
in the mathematical form of the prediction and filtering operators.  Moreover,
the Wiener-Kolmogorov framework forces us to sharply separate the linear and
nonlinear aspects of prediction and filtering, because it has a great deal of
trouble calculating nonlinear operators \cite{Wiener-nonlinear}.  Computational mechanics
is completely indifferent to this issue, since it packs {\it all} of the
process's structure into the $\epsilon$-machine, which is equally calculable in
linear or strongly nonlinear situations.

\subsection{Formal Language Theory and Grammatical Inference}

A formal language is a set of symbol strings (``words'' or ``allowed words'')
drawn from a finite alphabet.  Every formal language may be described either by
a set of rules (a ``grammar'') for creating all and only the allowed words, by
an abstract automaton which also generates the allowed words, or by an
automaton which accepts the allowed words and rejects all ``forbidden'' words.
Our $\epsilon$-machines, stripped of probabilities, correspond to such
automata---generative in the simple case or classificatory, if we add a reject
state and move to it when none of the allowed symbols are encountered.

Since Chomsky \cite{Chomsky-three-models,Chomsky-syntactic-structures}, it has been known that formal languages can
be classified into a hierarchy, the higher levels of which have strictly
greater expressive power.  The hierarchy is defined by restricting the form of
the grammatical rules or, equivalently, by limiting the amount and kind of
memory available to the automata.  The lowest level of the hierarchy is that of
regular languages, which may be familiar to Unix-using readers as regular
expressions.  These correspond to finite-state machines and to hidden Markov
models of finite dimension.  In such cases, relatives of our minimality and
uniqueness theorems are well known \cite{Hopcroft-Ullman}, and the construction of
causal states is analogous to the ``Nerode equivalence classing'' procedure
\cite{Hopcroft-Ullman,Trakhtenbrot-and-Barzdin}.  Our theorems, however, are {\it not} restricted to this
low-memory, non-stochastic setting.

The problem of learning a language from observational data has been extensively
studied by linguists, and by computer scientists interested in natural-language
processing.  Unfortunately, well developed learning techniques exist only for
the two lowest classes in the Chomsky hierarchy, the regular and the
context-free languages.  (For a good account of these procedures see
Ref.~\cite{Charniak}.)  Adapting and extending this work to the reconstruction
of $\epsilon$-machines should form a useful area of future research, a point to
which we alluded in the concluding remarks.

\subsection{Computational and Statistical Learning Theory}

The goal of computational learning theory \cite{Kearns-Vazirani,Vapnik-nature} is to identify
algorithms that quickly, reliably, and simply lead to good representations of a
target ``concept''.  The latter is typically defined to be a binary dichotomy
of a certain feature or input space.  Particular attention is paid to results
about ``probably approximately correct'' (PAC) procedures \cite{Valiant-learnable}: those
having a high probability of finding members of a fixed ``representation
class'' (e.g., neural nets, Boolean functions in disjunctive normal form,
and deterministic finite automata).  The key word here is ``fixed''; as in
contemporary time-series analysis, practitioners of this discipline acknowledge
the importance of getting the representation class right.  (Getting it wrong
can make easy problems intractable.)  In practice, however, they simply take
the representation class as a given, even assuming that we can always count on
it having at least one representation which {\it exactly} captures the target
concept.  Although this is in line with implicit assumptions in most of
mathematical statistics, it seems dubious when analyzing learning in the real
world \cite{Calculi-of-emergence,Boden-precis,Thornton-trash}.

In any case, the preceding development made no such assumption.  One of the
goals of computational mechanics is, exactly, {\it discovering} the best
representation.  This is not to say that the results of computational learning
theory are not remarkably useful and elegant, nor that one should not take
every possible advantage of them in implementing $\epsilon$-machine
reconstruction.  In our view, though, these theories belong more to statistical
inference, particularly to algorithmic parameter estimation, than to
foundational questions about the nature of pattern and the dynamics of
learning.

Finally, in a sense computational mechanics' focus on causal states is a
search for a particular kind of structural decomposition for a process. That
decomposition is most directly reflected in the conditional independence of
past and future that causal states induce. This decomposition reminds one
of the important role that conditional independence plays in contemporary
methods for artificial intelligence, both for developing systems that reason
in fluctuating environments \cite{Pearl-causality} and the more recently developed
algorithmic methods of graphical models \cite{Jordan-learning-in-graphical-models,Spirtes-Glymour-Scheines}.

\subsection{Description-Length Principles and Universal Coding Theory}

Rissanen's {\em minimum description length} (MDL) principle, most fully
described in Ref.~\cite{Rissanen-SCiSI}, is a procedure for selecting the most concise
generative model out of a family of models that are all statistically
consistent with given data.  The MDL approach starts from Shannon's results on
the connection between probability distributions and codes.  Rissanen's
development follows the inductive framework introduced by Solomonoff
\cite{Solomonoff}.

Suppose we choose a representation that leads to a class $\cal M$ of models and
are given data set $X$.  The MDL principle enjoins us to pick the model ${\rm
M} \in {\cal M}$ that minimizes the sum of the length of the description of $X$
given $\rm M$, plus the length of description of $\rm M$ given $\cal M$.  The
description length of $X$ is taken to be $-\log{\Prob(X|{\rm M})}$;
cf.~Eq.~(\ref{entropy-defined}).  The description length of ${\rm M}$ may be
regarded as either given by some coding scheme or, equivalently, by some
distribution over the members of ${\cal M}$.  (Despite the similarities to
model estimation in a Bayesian framework \cite{Lindley-Bayesian-stats}, Rissanen does not
interpret this distribution as a Bayesian prior or regard description length as
a measure of evidential support.)

The construction of causal states is somewhat similar to the states estimated
in Rissanen's {\em context} algorithm \cite{Rissanen-SCiSI,Rissanen-1984} (and to the
``vocabularies'' built by universal coding schemes, such as the popular
Lempel-Ziv algorithm \cite{Lempel-Ziv-complexity,Ziv-Lempel-universal-algorithm}).  Despite the similarities, there
are significant differences.  For a random source---for which there is a single
causal state---the context algorithm estimates a number of states that diverges
(at least logarithmically) with the length of the data stream, rather than
inferring a single state, as $\epsilon$-machine reconstruction would.
Moreover, we avoid any reference to encodings of rival models or to prior
distributions over them; $\Cmu(\AlternateStateSet)$ is not a description
length.

\subsection{Measure Complexity}

Ref.~\cite{Grassberger-1986} proposed that the appropriate measure of the complexity
of a process was the ``minimal average Shannon information needed'' for
optimal prediction. This {\em true measure complexity} was to be taken as
the Shannon entropy of the states used by some optimal predictor. The same
paper suggested that it could be approximated (from below) by the excess
entropy; there called the {\em effective measure complexity}, as noted in
Sec.~\ref{Bounds} above. This is a position closely allied to that of
computational mechanics, to Rissanen's MDL principle, and to the minimal
embeddings introduced by the ``geometry of a time series'' methods
\cite{Geometry-from-a-time-series} just described.

In contrast to computational mechanics, however, the key notion of ``optimal
prediction'' was left undefined, as were the nature and construction of the
states of the optimal predictor. In fact, the predictors used required knowing
the process's underlying equations of motion. Moreover, the statistical
complexity $\Cmu(\CausalStateSet)$ differs from the measure complexities
in that it is based on the well defined causal states, whose optimal predictive
powers are in turn precisely defined. Thus, computational mechanics is an
operational and constructive formalization of the insights expressed in
Ref.~\cite{Grassberger-1986}.

\subsection{Hierarchical Scaling Complexity}

Introduced in Ref.~\cite[ch.~9]{Badii-Politi}, this approach seeks, like
computational mechanics, to extend certain traditional ideas of statistical
physics.  In brief, the method is to construct a hierarchy of ${n}^{th}$-order
Markov models and examine the convergence of their predictions with the real
distribution of observables as $n\rightarrow\infty$.  The discrepancy between
prediction and reality is, moreover, defined information theoretically, in
terms of the relative entropy or Kullback-Leibler distance
\cite{Cover-and-Thomas,Kullback-info-theory-and-stats}.  (We have not used this quantity.)  The approach
implements Weiss's discovery that for finite-state sources there is a
structural distinction between block-Markovian sources ({\em subshifts of
finite type}) and {\em sofic systems}.  Weiss showed that, despite their finite
memory, sofic systems are the limit of an infinite series of increasingly
larger block-Markovian sources \cite{Weiss-1973}.

The hierarchical-scaling-complexity approach has several advantages,
particularly its ability to handle issues of scaling in a natural way
(see Ref.~\cite[sec.~9.5]{Badii-Politi}). Nonetheless, it does not attain all the
goals set in Sec.~\ref{Desiderata}. Its Markovian predictors are so many black
boxes, saying little or nothing about the hidden states of the process, their
causal connections, or the intrinsic computation carried on by the process.
All of these properties, as we have shown, are manifest from the
$\epsilon$-machine.  We suggest that a productive line of future work would be
to investigate the relationship between hierarchical scaling complexity and
computational mechanics, and to see whether they can be synthesized.  Along
these lines, hierarchical scaling complexity reminds us somewhat of
hierarchical $\epsilon$-machine reconstruction described in
Ref.~\cite{Calculi-of-emergence}.

\subsection{Continuous Dynamical Computing}

Using dynamical systems as computers has become increasingly attractive over
the last ten years or so among physicists, computer scientists, and others
exploring the physical basis of computation
\cite{Moore-recursion-theory,Moore-dynamical-recognizers,Orponen-survey,Blum-Shub-Smale}.  These proposals have ranged from highly
abstract ideas about how to embed Turing machines in discrete-time nonlinear
continuous maps \cite{Computation-at-the-onset,Moore-undecidability} to, more recently, schemes for
specialized numerical computation that could in principle be implemented in
current hardware \cite{Sinha-Ditto-prl}.  All of them, however, have been synthetic, in
the sense that they concern {\em designing} dynamical systems that implement
a given desired computation or family of computations. In contrast, one of the
central questions of computational mechanics is exactly the converse:
{\it given} a dynamical system, how can one detect what it is intrinsically
computing?

We believe that having a mathematical basis and a set of tools for answering
this question are important to the synthetic, engineering approach to dynamical
computing.  Using these tools we may be able to discover, for example, novel
forms of computation embedded in natural processes that operate at higher
speeds, with less energy, and with fewer physical degrees of freedom than
currently possible.

\end{appendix}

% **************************References**************************
\addcontentsline{toc}{section}{References}

\bibliography{locusts}

\begin{thebibliography}{100}

\bibitem{Yeomans}
Julia~M. Yeomans.
\newblock {\em Statistical Mechanics of Phase Transitions}.
\newblock Clarendon Press, Oxford, 1992.

\bibitem{Manneville-dissipative-structures}
Paul Manneville.
\newblock {\em Dissipative Structures and Weak Turbulence}.
\newblock Academic Press, Boston, Massachusetts, 1990.

\bibitem{Chaikin-Lubensky}
P.~M. Chaikin and T.~C. Lubensky.
\newblock {\em Principles of Condensed Matter Physics}.
\newblock Cambridge University Press, Cambridge, England, 1995.

\bibitem{Cross-Hohenberg}
Mark~C. Cross and Pierre Hohenberg.
\newblock Pattern {F}ormation {O}ut of {E}quilibrium.
\newblock {\em Reviews of Modern Physics}, 65:851--1112, 1993.

\bibitem{Calculi-of-emergence}
James~P. Crutchfield.
\newblock The calculi of emergence: {C}omputation, dynamics, and induction.
\newblock {\em Physica D}, 75:11--54, 1994.

\bibitem{Inferring-stat-compl}
James~P. Crutchfield and Karl Young.
\newblock Inferring statistical complexity.
\newblock {\em Physical Review Letters}, 63:105--108, 1989.

\bibitem{Computation-at-the-onset}
James~P. Crutchfield and Karl Young.
\newblock Computation at the onset of chaos.
\newblock In Zurek \cite{complexity-entropy-physics-of-info}, pages 223--269.

\bibitem{Perry-Binder-finite-stat-compl}
Nicol\'as Perry and P.-M. Binder.
\newblock Finite statistical complexity for sofic systems.
\newblock {\em Physical Review E}, 60:459--463, 1999.

\bibitem{Comp-mech-of-CA-example}
James~E. Hanson and James~P. Crutchfield.
\newblock Computational mechanics of cellular automata: {A}n example.
\newblock {\em Physica D}, 103:169--189, 1997.

\bibitem{Upper-thesis}
Daniel~R. Upper.
\newblock {\em Theory and Algorithms for Hidden {Markov} Models and Generalized
  Hidden {Markov} Models}.
\newblock PhD thesis, University of California, Berkeley, 1997.

\bibitem{JPC-MM-PNAS}
James~P. Crutchfield and Melanie Mitchell.
\newblock The evolution of emergent computation.
\newblock {\em Proceedings of the National Academy of Sciences},
  92:10742--10746, 1995.

\bibitem{Witt-et-al-1997}
A.~Witt, A.~Neiman, and J.~Kurths.
\newblock Characterizing the dynamics of stochastic bistable systems by
  measures of complexity.
\newblock {\em Physical Review E}, 55:5050--5059, 1997.

\bibitem{Delgado-collective-induced}
Jordi Delgado and Ricard~V. Sol{\'e}.
\newblock Collective-induced computation.
\newblock {\em Physical Review E}, 55:2338--2344, 1997.

\bibitem{Goncalves-dripping-faucet}
W.~M. Gon\c{c}alves, R.~D. Pinto, J.~C. Sartorelli, and M.~J. {de Oliveira}.
\newblock Inferring statistical complexity in the dripping faucet experiment.
\newblock {\em Physica A}, 257:385--389, 1998.

\bibitem{TDCS}
James~P. Crutchfield and Cosma~Rohilla Shalizi.
\newblock Thermodynamic depth of causal states: {O}bjective complexity via
  minimal representations.
\newblock {\em Physical Review E}, 59:275--283, 1999.

\bibitem{Borges-inquisitions}
Jorge~Luis Borges.
\newblock {\em Other {I}nquisitions, 1937--1952}.
\newblock University of Texas Press, Austin, 1964.
\newblock Trans. Ruth L. C. Simms.

\bibitem{JPC-semantics}
James~P. Crutchfield.
\newblock Semantics and thermodynamics.
\newblock In Martin Casdagli and Stephen Eubank, editors, {\em Nonlinear
  Modeling and Forecasting}, volume~12 of {\em Santa Fe Institute Studies in
  the Sciences of Complexity}, pages 317--359, Reading, Massachusetts, 1992.
  Addison-Wesley.

\bibitem{Phaedrus}
Plato.
\newblock {\em {Phaedrus}}.

\bibitem{Luria-working-brain}
A.~R. Luria.
\newblock {\em The {W}orking Brain: {A}n Introduction to Neuropsychology}.
\newblock Basic Books, New York, 1973.

\bibitem{Graham-pattern-analyzers}
Norma Van~Surdam Graham.
\newblock {\em Visual Pattern Analyzers}, volume~16 of {\em Oxford Psychology
  Series}.
\newblock Oxford University Press, Oxford, 1989.

\bibitem{Shettleworth}
Sara~J. Shettleworth.
\newblock {\em Cognition, Evolution and Behavior}.
\newblock Oxford University Press, Oxford, 1998.

\bibitem{Tou-and-Gonzalez}
Julius~T. Tou and Rafael~C. Gonzalez.
\newblock {\em Pattern Recognition Principles}.
\newblock Addison-Wesley, Reading, Mass, 1974.

\bibitem{Banks-processing}
Stephen~P. Banks.
\newblock {\em Signal Processing, Image Processing, and Pattern Recognition}.
\newblock Prentice Hall, New York, 1990.

\bibitem{Lim-two-d-processing}
Jae~S. Lim.
\newblock {\em Two-Dimensional Signal and Image Processing}.
\newblock Prentice Hall, New York, 1990.

\bibitem{Meno}
Plato.
\newblock {\it {Meno}}.
\newblock In Sec.~80D Meno says: ``How will you look for it, Socrates, when you
  do not know at all what it is? How will you aim to search for something you
  do not know at all? If you should meet it, how will you know that this is the
  thing that you did not know?'' The same difficulty is raised in {\it
  Theaetetus}, Sec.~197 {\it et seq.}

\bibitem{Principia}
Alfred~North Whitehead and Bertrand Russell.
\newblock {\em Principia Mathematica}.
\newblock Cambridge University Press, Cambridge, England, 2nd edition,
  1925--27.

\bibitem{Russell-intro-to-math-phil}
Bertrand Russell.
\newblock {\em Introduction to Mathematical Philosophy}.
\newblock The Muirhead Library of Philosophy. George Allen and Unwin, London,
  revised edition, 1920.
\newblock First edition, 1919. Reprinted New York: Dover Books, 1993.

\bibitem{JPC-information-and-its-metric}
James~P. Crutchfield.
\newblock Information and its metric.
\newblock In L.~Lam and H.~C. Morris, editors, {\em Nonlinear Structures in
  Physical Systems---Pattern Formation, Chaos and Waves}, page 119, New York,
  1990. Springer-Verlag.

\bibitem{Russell-human-knowledge}
Bertrand Russell.
\newblock {\em Human Knowledge: {Its} Scope and Limits}.
\newblock Simon and Schuster, New York, 1948.

\bibitem{Rhodes-book}
John Rhodes.
\newblock {\em Applications of Automata Theory and Algebra via the Mathematical
  Theory of Complexity to Biology, Physics, Psychology, Philosophy, Games, and
  Codes}.
\newblock University of California, Berkeley, California, 1971.

\bibitem{Nehaniv-Rhodes-evolution}
Chrystopher~L. Nehaniv and John~L. Rhodes.
\newblock Krohn-{R}hodes theory, hierarchies, and evolution.
\newblock In Boris Mirkin, F.~R. McMorris, Fred~S. Roberts, and Andrey
  Rzhetsky, editors, {\em Mathematical Hierarchies and Biology: DIMACS
  workshop, November 13--15, 1996}, volume~37 of {\em DIMACS Series in Discrete
  Mathematics and Theoretical Computer Science}, Providence, Rhode Island,
  1997. American Mathematical Society.

\bibitem{Grenander-elements}
Ulf Grenander.
\newblock {\em Elements of Pattern Theory}.
\newblock Johns Hopkins Studies in the Mathematical Sciences. Johns Hopkins
  University Press, Baltimore, Maryland, 1996.

\bibitem{Grenander-hands}
Ulf Grenander, Y.~Chow, and D.~M. Keenan.
\newblock {\em Hands: {A} Pattern Theoretic Study of Biological Shapes},
  volume~2 of {\em Research Notes in Neural Computing}.
\newblock Springer-Verlag, New York, 1991.

\bibitem{Grenander-potatoes}
Ulf Grenander and K.~Manbeck.
\newblock A stochastic shape and color model for defect detection in potatoes.
\newblock {\em American Statistical Association}, 2:131--151, 1993.

\bibitem{Kolmogorov-three-approaches}
A.~N. Kolmogorov.
\newblock Three approaches to the quantitative definition of information.
\newblock {\em Problems of Information Transmission}, 1:1--7, 1965.

\bibitem{Chaitin-KC-complexity-cite}
Gregory Chaitin.
\newblock On the length of programs for computing finite binary sequences.
\newblock {\em Journal of the Association for Computing Machinery},
  13:547--569, 1966.

\bibitem{Kolmogorov-1983}
A.~N. Kolmogorov.
\newblock Combinatorial foundations of information theory and the calculus of
  probabilities.
\newblock {\em Russ. Math. Surveys}, 38:29, 1983.

\bibitem{Li-and-Vitanyi-1993}
Ming Li and Paul M.~B. Vitanyi.
\newblock {\em An Introduction to {Kolmogorov} Complexity and its
  Applications}.
\newblock Springer-Verlag, New York, 1993.

\bibitem{Minsky-computation}
Marvin Minsky.
\newblock {\em Computation: {F}inite and Infinite Machines}.
\newblock Prentice-Hall, Englewood Cliffs, New Jersey, 1967.

\bibitem{Martin-Lof}
P.~Martin-{L\"of}.
\newblock The definition of random sequences.
\newblock {\em Information and Control}, 9:602--619, 1966.

\bibitem{Levin-information-conservation}
L.~A. Levin.
\newblock Laws of information conservation (nongrowth) and aspects of the
  foundation of probability theory.
\newblock {\em Problemy Peredachi Informatsii}, 10:30--35, 1974.
\newblock Translation: {\it Problems of Information Transmission} {\bf 10}
  (1974) 206--210.

\bibitem{Gurzadyan}
V.~G. Gurzadyan.
\newblock Kolmogorov complexity as a descriptor of cosmic microwave background
  maps.
\newblock {\em Europhysics Letters}, 46:114--117, 1999.
\newblock Also available as an electronic preprint, LANL archive,
  astro-phy/9902123.

\bibitem{Solomonoff}
Raymond~J. Solomonoff.
\newblock A formal theory of inductive inference.
\newblock {\em Information and Control}, 7:1--22 and 224--254, 1964.

\bibitem{Vitanyi-and-Li-1999}
Paul Vit{\'a}nyi and Ming Li.
\newblock Minimum description length induction, {Bayesianism}, and {Kolmogorov}
  complexity.
\newblock Electronic pre-print, LANL Archive, cs.LG/9901014, 1999.

\bibitem{Flake}
Gary~William Flake.
\newblock {\em The Computational Beauty of Nature: {C}omputer Explorations of
  Fractals, Chaos, Complex Systems and Adaptation}.
\newblock MIT Press, Cambridge, Massachusetts, 1998.

\bibitem{Rissanen-SCiSI}
Jorma Rissanen.
\newblock {\em Stochastic Complexity in Statistical Inquiry}.
\newblock World Scientific, Singapore, 1989.

\bibitem{Bennett-how-and-why}
Charles~H. Bennett.
\newblock How to define complexity in physics, and why.
\newblock In Zurek \cite{complexity-entropy-physics-of-info}, pages 137--148.

\bibitem{Koppel-1987}
Moshe Koppel.
\newblock Complexity, depth, and sophistication.
\newblock {\em Complex Systems}, 1:1087--1091, 1987.

\bibitem{Koppel-Atlan}
Moshe Koppel and Henri Atlan.
\newblock An almost machine-independent theory of program-length complexity,
  sophistication and induction.
\newblock {\em Information Sciences}, 56:23--44, 1991.

\bibitem{Dennett-real-patterns}
Daniel~C. Dennett.
\newblock Real patterns.
\newblock {\em Journal of Philosophy}, 88:27--51, 1991.
\newblock Reprinted in Dennett (1997).

\bibitem{Anything-ever-new}
James~P. Crutchfield.
\newblock Is anything ever new? {C}onsidering emergence.
\newblock In G.~Cowan, D.~Pines, and D.~Melzner, editors, {\em Complexity:
  {M}etaphors, Models, and Reality}, volume~19 of {\em Santa Fe Institute
  Studies in the Sciences of Complexity}, pages 479--497, Reading,
  Massachusetts, 1994. Addison-Wesley.

\bibitem{Holland-emergence}
John~H. Holland.
\newblock {\em Emergence: {F}rom Chaos to Order}.
\newblock Addison-Wesley, Reading, Massachusetts, 1998.

\bibitem{Boltzmann-gas-theory}
Ludwig Boltzmann.
\newblock {\em Lectures on Gas Theory}.
\newblock University of California Press, Berkeley, 1964.

\bibitem{Cramer}
Harald Cram{\'e}r.
\newblock {\em Mathematical Methods of Statistics}.
\newblock Almqvist and Wiksells, Uppsala, 1945.
\newblock Republished by Princeton University Press, 1946, as vol. 9 in the
  Princeton Mathematics Series, and as a paperback, in the Princeton Landmarks
  in Mathematics and Physics series, 1999.

\bibitem{Shannon-1948}
Claude~E. Shannon.
\newblock A mathematical theory of communication.
\newblock {\em Bell System Technical Journal}, 27:379--423, 1948.

\bibitem{Hume-treatise}
David Hume.
\newblock {\em A {T}reatise of {H}uman {N}ature: {B}eing an {A}ttempt to
  {I}ntroduce the {E}xperimental {M}ethod of {R}easoning into {M}oral
  {S}ubjects}.
\newblock John Noon, London, 1739.
\newblock Reprint (Oxford: Clarendon Press, 1951) of original edition, with
  notes and analytical index.

\bibitem{Bunge}
Mario Bunge.
\newblock {\em Causality: {T}he Place of the Causal Princple in Modern
  Science}.
\newblock Harvard University Press, Cambridge, Massachusetts, 1959.
\newblock Reprinted as {\it Causality and Modern Science}, NY: Dover Books,
  1979.

\bibitem{Salmon-1984}
Wesley~C. Salmon.
\newblock {\em Scientific Explanation and the Causal Structure of the World}.
\newblock Princeton University Press, Princeton, 1984.

\bibitem{Billingsley-ergodic-theory-and-info}
Patrick Billingsley.
\newblock {\em Ergodic Theory and Information}.
\newblock Tracts on Probablity and Mathematical Statistics. Wiley, New York,
  1965.

\bibitem{Billingsley-probability-and-measure}
Patrick Billingsley.
\newblock {\em Probability and Measure}.
\newblock Wiley Series in Probability and Mathematical Statistics. Wiley, New
  York, 1979.

\bibitem{Schutz-geometrical}
Bernard~F. Schutz.
\newblock {\em Geometrical Methods of Mathematical Physics}.
\newblock Cambridge University Press, Cambridge, England, 1980.

\bibitem{Cover-and-Thomas}
Thomas~M. Cover and Joy~A. Thomas.
\newblock {\em Elements of Information Theory}.
\newblock Wiley, New York, 1991.

\bibitem{Occam}
William~of Ockham.
\newblock {\em Philosophical Writings: {A} Selection, Translated, with an
  Introduction, by Philotheus Boehner, O.F.M., Late Professor of Philosophy,
  The Franciscan Institute}.
\newblock Bobbs-Merrill, Indianapolis, 1964.
\newblock first pub. various European cities, early 1300s.

\bibitem{Kuan-Yin-Tzu}
Anonymous.
\newblock {K}uan {Y}in {T}zu, T'ang Dynasty.
\newblock Written in China during the T'ang dynasty. Partial translation in
  Joseph Needham, {\it Science and Civilisation in China}, vol. II (Cambridge
  University Press, 1956), p. 73.

\bibitem{DNCO}
David~P. Feldman and James~P. Crutchfield.
\newblock Discovering non-critical organization: {S}tatistical mechanical,
  information theoretic, and computational views of patterns in simple
  one-dimensional spin systems.
\newblock {\em Journal of Statistical Physics}, submitted, 1998.
\newblock Santa Fe Institute Working Paper 98-04-026,
  http://www.santafe.edu/projects/CompMech/ papers/ DNCO.html.

\bibitem{Hopcroft-Ullman}
John~E. Hopcroft and Jeffrey~D. Ullman.
\newblock {\em Introduction to Automata Theory, Languages, and Computation}.
\newblock Addison-Wesley, Reading, 1979.
\newblock 2nd edition of {\it Formal Languages and Their Relation to Automata},
  1969.

\bibitem{Kemeny-finite-chains}
John~G. Kemeny and J.~Laurie Snell.
\newblock {\em Finite {Markov} Chains}.
\newblock Springer-Verlag, New York, 1976.

\bibitem{Kemeny-denumerable-chains}
John~G. Kemeny, J.~Laurie Snell, and Anthony~W. Knapp.
\newblock {\em Denumerable {Markov} Chains}.
\newblock Springer-Verlag, New York, 2nd edition, 1976.

\bibitem{Hanson-thesis}
James~E. Hanson.
\newblock {\em Computational Mechanics of Cellular Automata}.
\newblock PhD thesis, University of California, Berkeley, 1993.

\bibitem{Bateson-mind-and-nature}
Gregory Bateson.
\newblock {\em Mind and Nature: {A} Necessary Unity}.
\newblock E. P. Dutton, New York, 1979.

\bibitem{Kullback-info-theory-and-stats}
Solomon Kullback.
\newblock {\em Information Theory and Statistics}.
\newblock Dover Books, New York, 2nd edition, 1968.
\newblock First edition New York: Wiley, 1959.

\bibitem{Bernard}
Claude Bernard.
\newblock {\em Introduction a l'etude de la medecine experimentale}.
\newblock J. B. Bailliere, Paris, 1865.
\newblock Trans. by Henry Copley Green as {\it Introduction to the Study of
  Experimental Medicine}, New York: Macmillian, 1927; reprinted New York:
  Dover, 1957.

\bibitem{JPC-Packard-noisy-chaos}
James~P. Crutchfield and Norman~H. Packard.
\newblock Symbolic dynamics of noisy chaos.
\newblock {\em Physica D}, 7:201--223, 1983.

\bibitem{Shaw-dripping}
Robert Shaw.
\newblock {\em The Dripping Faucet as a Model Chaotic System}.
\newblock Aerial Press, Santa Cruz, California, 1984.

\bibitem{Grassberger-1986}
Peter Grassberger.
\newblock Toward a quantitative theory of self-generated complexity.
\newblock {\em International Journal of Theoretical Physics}, 25:907--938,
  1986.

\bibitem{Lindgren-Nordahl-1988}
Kristian Lindgren and Mats~G. Nordahl.
\newblock Complexity measures and cellular automata.
\newblock {\em Complex Systems}, 2:409--440, 1988.

\bibitem{Li-complexity-vs-entropy}
W.~Li.
\newblock On the relationship between complexity and entropy for {M}arkov
  chains and regular languages.
\newblock {\em Complex Systems}, 5:381--399, 1991.

\bibitem{Arnold-info-theory-phase-trans}
Dirk Arnold.
\newblock Information-theoretic analysis of phase transitions.
\newblock {\em Complex Systems}, 10:143--155, 1996.

\bibitem{Bialek-predictive-info}
William Bialek and Naftali Tishby.
\newblock Predictive information.
\newblock Electronic pre-print, LANL archive, cond-mat/9902341, 1999.

\bibitem{JPC-DPF-stat-compl-of-1d-spin}
James~P. Crutchfield and David~P. Feldman.
\newblock Statistical complexity of simple one-dimensional spin systems.
\newblock {\em Physical Review E}, 55:1239R--1243R, 1997.

\bibitem{Ashby-I-to-C}
W.~Ross Ashby.
\newblock {\em An {I}ntroduction to {C}ybernetics}.
\newblock Chapman and Hall, London, 1956.

\bibitem{Touchette-Lloyd}
Hugo Touchette and Seth Lloyd.
\newblock Information-theoretic limits of control.
\newblock {\em Physical Review Letters}, 84:1156--1159, 1999.

\bibitem{Lempel-Ziv-two-d}
Abraham Lempel and Jacob Ziv.
\newblock Compression of two-dimensional data.
\newblock {\em IEEE Transactions in Information Theory}, IT-32:2--8, 1986.

\bibitem{DPF-thesis}
David~P. Feldman.
\newblock {\em Computational Mechanics of Classical Spin Systems}.
\newblock PhD thesis, University of California, Davis, 1998.
\newblock Available on-line at http://hornacek.coa.edu/dave/Thesis/thesis.html.

\bibitem{Mayo-error}
Deborah Mayo.
\newblock {\em Error and the Growth of Experimental Knowledge}.
\newblock Science and Its Conceptual Foundations. University of Chicago Press,
  Chicago, 1996.

\bibitem{JPC-and-Chris-Douglas}
James~P. Crutchfield and Cristopher Douglas.
\newblock Imagined complexity: Learning a random process.
\newblock in preparation, 1999.

\bibitem{Lidl}
Rudolf Lidl and Gunter Pilz.
\newblock {\em Applied Abstract Algebra}.
\newblock Springer, New York, 1984.

\bibitem{Ljapin-semigroups}
E.~S. Ljapin.
\newblock {\em Semigroups}, volume~3 of {\em Translations of Mathematical
  Monographs}.
\newblock American Mathematical Society, Providence, Rhode Island, 1963.

\bibitem{Young-thesis}
Karl Young.
\newblock {\em The Grammar and Statistical Mechanics of Complex Physical
  Systems}.
\newblock PhD thesis, University of California, Santa Cruz, 1991.

\bibitem{Geometry-from-a-time-series}
Norman~H. Packard, James~P. Crutchfield, J.~Doyne Farmer, and Robert~S. Shaw.
\newblock Geometry from a time series.
\newblock {\em Physical Review Letters}, 45:712--716, 1980.

\bibitem{Takens-embedding}
Floris Takens.
\newblock Detecting strange attractors in fluid turbulence.
\newblock In D.~A. Rand and L.~S. Young, editors, {\em Symposium on Dynamical
  Systems and Turbulence}, volume 898 of {\em Lecture Notes in Mathematics},
  page 366, Berlin, 1981. Springer-Verlag.

\bibitem{JPC-equations-of-motion}
James~P. Crutchfield and Bruce~S. McNamara.
\newblock Equations of motion from a data series.
\newblock {\em Complex Systems}, 1:417--452, 1987.

\bibitem{Neymman-first-course}
Jerzy Neyman.
\newblock {\em First Course in Probability and Statistics}.
\newblock Henry Holt, New York, 1950.

\bibitem{Blackwell-Girshick}
David Blackwell and M.~A. Girshick.
\newblock {\em Theory of Games and Statistical Decisions}.
\newblock Wiley, New York, 1954.
\newblock Reprinted New York: Dover Books, 1979.

\bibitem{Luce-and-Raiffa}
R.~Duncan Luce and Howard Raiffa.
\newblock {\em Games and Decisions: {I}ntroduction and Critical Survey}.
\newblock Wiley, New York, 1957.

\bibitem{Gelfand-Yaglom-info}
I.~M. Gel'fand and A.~M. Yaglom.
\newblock Calculation of the amount of information about a random function
  contained in another such function.
\newblock {\em Uspekhi Matematicheski Nauk}, 12:3--52, 1956.
\newblock Trans. in {\it American Mathematical Society Translations}, 2nd
  series, {\bf 12} (1959): 199--246.

\bibitem{Caines}
Peter~E. Caines.
\newblock {\em Linear Stochastic Systems}.
\newblock Wiley, New York, 1988.

\bibitem{Gray-entropy}
Robert~M. Gray.
\newblock {\em Entropy and Information Theory}.
\newblock Springer-Verlag, New York, 1990.

\bibitem{Blackwell-identifiability}
David Blackwell and Lambert Koopmans.
\newblock On the identifiability problem for functions of finite {Markov}
  chains.
\newblock {\em Annals of Mathematical Statistics}, 28:1011--1015, 1957.

\bibitem{Ito-et-al-identifiability}
H.~Ito, S.-I. Amari, and K.~Kobayashi.
\newblock Identifiability of hidden {Markov} information sources and their
  minimum degrees of freedom.
\newblock {\em IEEE Transactions on Information Theory}, 38:324--333, 1992.

\bibitem{Jaeger-operator-models}
H.~Jaeger.
\newblock Observable operator models for discrete stochastic time series.
\newblock {\em Neural Computation}, forthcoming, 1999.
\newblock ftp://ftp.gmd.de/GMD/ais/ publications/1999/.

\bibitem{Algoet-universal-schemes}
Paul Algoet.
\newblock Universal schemes for prediction, gambling and portfolio selection.
\newblock {\em The Annals of Probability}, 20:901--941, 1992.
\newblock See also an important Correction, {\it The Annals of Probability},
  {\bf 23} (1995): 474--478.

\bibitem{Kolmogorov-interpolation-extrapolation}
A.~N. Kolmogorov.
\newblock Interpolation und extrapolation von station{\"a}ren zuf{\"a}lligen
  folgen.
\newblock {\em Bull. Acad. Sci. U.S.S.R., Math.}, 3:3--14, 1941.
\newblock In German.

\bibitem{Wiener-time-series}
Norbert Wiener.
\newblock {\em Extrapolation, Interpolation, and Smoothing of Stationary Time
  Series: {W}ith Engineering Applications}.
\newblock The Technology Press of the Massachusetts Institute of Technology,
  Cambridge, Massachusetts, 1949.
\newblock ``First published during the war as a classifed report to Section
  ${\rm D}_{2}$, National Defense Research Council''.

\bibitem{Wiener-nonlinear}
Norbert Wiener.
\newblock {\em Nonlinear Problems in Random Theory}.
\newblock The Technology Press of the Massachusetts Institute of Technology,
  Cambridge, Massachusetts, 1958.

\bibitem{Wiener-cybernetics}
Norbert Wiener.
\newblock {\em Cybernetics: {O}r, Control and Communication in the Animal and
  the Machine}.
\newblock MIT Press, Cambridge, Massachusetts, 2nd edition, 1961.
\newblock First edition New York: Wiley, 1948.

\bibitem{Chomsky-three-models}
Noam Chomsky.
\newblock Three models for the description of language.
\newblock {\em IRE Transactions on Information Theory}, 2:113, 1956.

\bibitem{Chomsky-syntactic-structures}
Noam Chomsky.
\newblock {\em Syntactic Structures}, volume~4 of {\em Janua linguarum, series
  minor}.
\newblock Mouton, The Hauge, 1957.

\bibitem{Trakhtenbrot-and-Barzdin}
B.~A. Trakhtenbrot and Ya.~M. Barzdin.
\newblock {\em Finite Automata}.
\newblock North-Holland, Amsterdam, 1973.

\bibitem{Charniak}
Eugene Charniak.
\newblock {\em Statistical Language Learning}.
\newblock Language, Speech and Communication. MIT Press, Cambridge,
  Massachusetts, 1993.

\bibitem{Kearns-Vazirani}
Michael~J. Kearns and Umesh~V. Vazirani.
\newblock {\em An Introduction to Computational Learning Theory}.
\newblock MIT Press, Cambridge, Massachusetts, 1994.

\bibitem{Vapnik-nature}
V.~N. Vapnik.
\newblock {\em The Nature of Statistical Learning Theory}.
\newblock Springer-Verlag, Berlin, 2nd edition, 2000.

\bibitem{Valiant-learnable}
Leslie~G. Valiant.
\newblock A theory of the learnable.
\newblock {\em Communications of the Association for Computing Machinery},
  27:1134--1142, 1984.

\bibitem{Boden-precis}
Margaret~A. Boden.
\newblock Precis of {{\it The}} {{\it Creative}} {{\it Mind:}} {{\it Myths}}
  {\it and} {{\it Mechanisms}}.
\newblock {\em Behaviorial and Brain Sciences}, 17:519--531, 1994.

\bibitem{Thornton-trash}
Chris Thornton.
\newblock {\em Truth from Trash: How Learning Makes Sense}.
\newblock Complex Adaptive Systems. MIT Press, Cambridge, Massachusetts, 2000.

\bibitem{Pearl-causality}
Judea Pearl.
\newblock {\em Causality: Models, Reasoning, and Inference}.
\newblock Cambridge University Press, Cambridge, England, 2000.

\bibitem{Jordan-learning-in-graphical-models}
M.~I. Jordan, editor.
\newblock {\em Learning in Graphical Models}, volume~89 of {\em NATO Science
  Series D: Behavioral and Social Sciences}, Dordrecht, 1998.

\bibitem{Spirtes-Glymour-Scheines}
Peter Spirtes, Clark Glymour, and Richard Scheines.
\newblock {\em Causation, Prediction, and Search}.
\newblock Adaptive Computation and Machine Learning. MIT Press, Cambridge,
  Massachusetts, 2000.

\bibitem{Lindley-Bayesian-stats}
David~V. Lindley.
\newblock {\em Bayesian Statistics, a Review}.
\newblock Society for Industrial and Applied Mathematics, Philadelphia, 1972.

\bibitem{Rissanen-1984}
Jorma Rissanen.
\newblock Universal coding, information, prediction, and estimation.
\newblock {\em IEEE Transactions in Information Theory}, IT-30:629--636, 1984.

\bibitem{Lempel-Ziv-complexity}
Abraham Lempel and Jacob Ziv.
\newblock On the complexity of finite sequences.
\newblock {\em IEEE Transactions in Information Theory}, IT-22:75--81, 1976.

\bibitem{Ziv-Lempel-universal-algorithm}
Jacob Ziv and Abraham Lempel.
\newblock A universal algorithm for sequential data compression.
\newblock {\em IEEE Transactions in Information Theory}, IT-23:337--343, 1977.

\bibitem{Badii-Politi}
Remo Badii and Antonio Politi.
\newblock {\em Complexity: {H}ierarchical Structures and Scaling in Physics},
  volume~6 of {\em Cambridge Nonlinear Science Series}.
\newblock Cambridge University Press, Cambridge, 1997.

\bibitem{Weiss-1973}
Benjamin Weiss.
\newblock Subshifts of finite type and sofic systems.
\newblock {\em Monatshefte f{\"{u}}r Mathematik}, 77:462--474, 1973.

\bibitem{Moore-recursion-theory}
Cristopher Moore.
\newblock Recursion theory on the reals and continuous-time computation.
\newblock {\em Theoretical Computer Science}, 162:23--44, 1996.

\bibitem{Moore-dynamical-recognizers}
Cristopher Moore.
\newblock Dynamical recognizers: Real-time language recognition by analog
  computers.
\newblock {\em Theoretical Computer Science}, 201:99--136, 1998.

\bibitem{Orponen-survey}
Pekka Orponen.
\newblock A survey of continuous-time computation theory.
\newblock In D.-Z. Du and K.-I Ko, editors, {\em Advances in Algorithms,
  Languages, and Complexity}, pages 209--224. Kluwer Academic, Dordrecht, 1997.

\bibitem{Blum-Shub-Smale}
Lenore Blum, Michael Shub, and Steven Smale.
\newblock On a theory of computation and complexity over the real numbers:
  {NP}-completeness, recursive functions and universal machines.
\newblock {\em Bulletin of the American Mathematical Society}, 21:1--46, 1989.

\bibitem{Moore-undecidability}
Cristopher Moore.
\newblock Unpredictability and undecidability in dynamical systems.
\newblock {\em Physical Review Letters}, 64:2354--2357, 1990.

\bibitem{Sinha-Ditto-prl}
Sudeshna Sinha and William~L. Ditto.
\newblock Dynamics based computation.
\newblock {\em Physical Review Letters}, 81:2156--2159, 1998.

\bibitem{complexity-entropy-physics-of-info}
Wojciech~H. Zurek, editor.
\newblock {\em Complexity, Entropy, and the Physics of Information}, volume~8
  of {\em Santa Fe Institute Studies in the Sciences of Complexity}, Reading,
  Massachusetts, 1990. Addison-Wesley.

\end{thebibliography}

% ***************************Glossary***************************
\onecolumn

\section*{Glossary of Notation}
\addcontentsline{toc}{section}{Glossary of Notation}

In the order of their introduction.

\hspace{10mm}

\begin{tabular}{cll}
Symbol & Description & Where Introduced \\

${\cal O}$ &
	Object in which we wish to find a pattern &
	Sec.~\ref{Patterns}, p.~\pageref{Patterns}\\

${\cal P}$ &
	Pattern in ${\cal O}$ &
	Sec.~\ref{Patterns}, p.~\pageref{Patterns}\\

${\cal A}$ &
	Countable alphabet &
	Sec.~\ref{HiddenProcesses}, p.~\pageref{HiddenProcesses}\\

$\BiInfinity$ &
	Bi-infinite, stationary, discrete stochastic process on ${\cal A}$ &
	Def.~\ref{AProcess}, p.~\pageref{AProcess} \\

$\biinfinity$ &
	Particular realization of $\BiInfinity$ &
	Def.~\ref{AProcess}, p.~\pageref{AProcess} \\

$\FutureL$ &
	Random variable for the next $L$ values of $\BiInfinity$ &
	Sec.~\ref{HiddenProcesses}, p.~\pageref{AProcess} \\

$\futureL$ &
	Particular value of $\FutureL$ &
	Sec.~\ref{HiddenProcesses}, p.~\pageref{AProcess} \\

$\NextObservable$ &
	Next observable generated by $\BiInfinity$ &
	Sec.~\ref{HiddenProcesses}, p.~\pageref{AProcess} \\

$\PastL$ &
	As $\FutureL$, but for the last $L$ values, up to the present &
	Sec.~\ref{HiddenProcesses}, p.~\pageref{AProcess} \\

$\pastL$ &
	Particular value of $\PastL$ &
	Sec.~\ref{HiddenProcesses}, p.~\pageref{AProcess} \\

$\Future$ &
	Semi-infinite future half of $\BiInfinity$ &
	Sec.~\ref{HiddenProcesses}, p.~\pageref{AProcess} \\

$\future$ &
	Particular value of $\Future$ &
	Sec.~\ref{HiddenProcesses}, p.~\pageref{AProcess} \\

$\Past$ &
	Semi-infinite past half of $\BiInfinity$ &
	Sec.~\ref{HiddenProcesses}, p.~\pageref{AProcess} \\

$\past$ &
	Particular value of $\Past$ &
	Sec.~\ref{HiddenProcesses}, p.~\pageref{AProcess} \\

$\lambda$ &
	Null string or null symbol &
	Sec.~\ref{HiddenProcesses}, p.~\pageref{AProcess} \\

$\AllPasts$ &
	Set of all pasts realized by the process $\BiInfinity$ &
	Sec.~\ref{ThePool}, p.~\pageref{ThePool}\\

$\AlternateStateSet$ &
	Partition of $\AllPasts$ into effective states &
	Sec.~\ref{ThePool}, p.~\pageref{ThePool}\\

$\alternatestate$ &
	Member-class of $\AlternateStateSet$; a particular effective state &
	Sec.~\ref{ThePool}, p.~\pageref{ThePool}\\

$\eta$ &
	Function from $\AllPasts$ to $\AlternateStateSet$ &
	Sec.~\ref{ThePool}, Eq.~(\ref{EtaDefn}), p.~\pageref{EtaDefn}\\

$\AlternateState$ &
	Current effective ($\eta$) state, as a random variable &
	Sec.~\ref{ThePool}, p.~\pageref{ThePool}\\

${\AlternateState}^{\prime}$ &
	Next effective state, as a random variable &
	Sec.~\ref{ThePool}, p.~\pageref{ThePool}\\

$H[X]$ &
	Entropy of the random variable $X$ &
	Sec.~\ref{EntropyDefn}, p.~\pageref{EntropyDefn}\\

$H[X, Y]$ &
	Joint entropy of the random variables $X$ and $Y$ &
	Sec.~\ref{JointCondEntropyDefn}, p.~\pageref{JointCondEntropyDefn}\\

$H[X|Y]$ &
	Entropy of $X$ conditioned on $Y$ &
	Sec.~\ref{JointCondEntropyDefn}, p.~\pageref{JointCondEntropyDefn}\\

$I[X;Y]$ &
	Mutual information of $X$ and $Y$ &
	Sec.~\ref{MutualInfoDefn}, p.~\pageref{MutualInfoDefn}\\

$\hmu[\Future]$ &
	Entropy rate of $\Future$ &
	Sec.~\ref{PatternsInEnsembles}, Eq.~(\ref{EntropyRateBlockDefn}), p.~\pageref{EntropyRateBlockDefn}\\

$\hmu[\Future|X]$ &
	Entropy rate of $\Future$ conditioned on $X$ &
	Sec.~\ref{PatternsInEnsembles}, Eq.~(\ref{ConditionalEntropyRateDefn}), p.~\pageref{ConditionalEntropyRateDefn}\\

$\Cmu(\AlternateStateSet)$ &
	Statistical complexity of $\AlternateStateSet$ &
	Def.~\ref{statistical-complexity-defined}, p.~\pageref{statistical-complexity-defined}\\

$\CausalStateSet$ &
	Set of the causal states of $\BiInfinity$ &
	Def.~\ref{CausalStatesFunctionDefn}, p.~\pageref{CausalStatesFunctionDefn}\\

$\causalstate$ &
	Particular causal state &
	Def.~\ref{CausalStatesFunctionDefn}, p.~\pageref{CausalStatesFunctionDefn}\\

$\epsilon$ &
	Function from histories to causal states &
	Def.~\ref{CausalStatesFunctionDefn}, p.~\pageref{CausalStatesFunctionDefn}\\

$\CausalState$ &
	Current causal state, as a random variable &
	Def.~\ref{CausalStatesFunctionDefn}, p.~\pageref{CausalStatesFunctionDefn}\\

${\CausalState}^{\prime}$ &
	Next causal state, as a random variable &
	Def.~\ref{CausalStatesFunctionDefn}, p.~\pageref{CausalStatesFunctionDefn}\\

$\CausalEquivalence$ &
	Relation of causal equivalence between two histories &
	Sec.~\ref{DefnCausalStatesEMs}, p.~\pageref{CausalStatesRelationDefn}\\

${T}^{(s)}_{ij}$ &
	Probability of going from causal state $i$ to $j$, emitting $s$ &
	Def.~\ref{CausalTransitionsDefn}, p.~\pageref{CausalTransitionsDefn}\\

$\PrescientStateSet$ &
	Set of prescient rival states &
	Def.~\ref{PrescientRivals}, p.~\pageref{PrescientRivals}\\

$\prescientstate$ &
	Particular prescient rival state &
	Def.~\ref{PrescientRivals}, p.~\pageref{PrescientRivals}\\

$\PrescientState$ &
	Current prescient rival state, as a random variable &
	Def.~\ref{PrescientRivals}, p.~\pageref{PrescientRivals}\\

${\PrescientState}^{\prime}$ &
	Next prescient rival state, as a random variable &
	Def.~\ref{PrescientRivals}, p.~\pageref{PrescientRivals}\\

$\Cmu({\cal O})$ &
	Statistical complexity of the process ${\cal O}$ &
	Def.~\ref{statistical-complexity-of-a-process}, p.~\pageref{statistical-complexity-of-a-process}\\

$\Cmu$ &
	Without an argument, short for $\Cmu({\cal O})$ &
	Def.~\ref{statistical-complexity-of-a-process}, p.~\pageref{statistical-complexity-of-a-process}\\

$\EE$ &
	Excess entropy &
	Def.~\ref{ExcessEntropyDefn}, p.~\pageref{ExcessEntropyDefn}\\

\end{tabular}

\twocolumn

\end{document}